\documentclass[superscriptaddress,aps,prd,nofootinbib,twocolumn,showpacs]{revtex4-1}

\usepackage[utf8]{inputenc}
\usepackage[british,UKenglish,USenglish,american]{babel}
\usepackage{amsmath}

\usepackage[draft]{graphicx}

\usepackage{float}

\usepackage{slashed,ulem}

\usepackage{mathrsfs}
\usepackage{amssymb}
\usepackage{bbm}

\usepackage{nicefrac}

\usepackage{microtype}

\usepackage{physics}

\usepackage{xcolor}

\usepackage{dsfont}

\usepackage{enumerate}

\newcommand\numberthis{\addtocounter{equation}{1}\tag{\theequation}}

\renewcommand{\bar}{\overline}

\usepackage{bm}

\newcommand{\e}{\mathrm{e}}

\newcommand{\comment}[1]{}

\begin{document}

\title{Neutron stars with large quark cores}

\author{Márcio Ferreira}
\email{marcio.ferreira@uc.pt}
\affiliation{CFisUC, Department of Physics, University of Coimbra, 
	P-3004 - 516  Coimbra, Portugal}

\author{Renan Câmara Pereira}
\email{renan.pereira@student.uc.pt}
\affiliation{CFisUC, Department of Physics, University of Coimbra, 
	P-3004 - 516  Coimbra, Portugal}

\author{Constança Providência}
\email{cp@uc.pt}
\affiliation{CFisUC, Department of Physics, University of Coimbra, 
	P-3004 - 516  Coimbra, Portugal}

\date{\today}

\begin{abstract}
	
	We describe charge-neutral neutron star matter in $\beta-$equilibrium using hybrid equations of state, where
	a first-order phase transition from hadronic to quark matter is realized.
	The hadronic matter is described in a model-independent way
	by a Taylor expansion around saturation density $n_0$, while the three-flavor NJL model is used for the quark matter.
	Exploring the present uncertainty on the empirical parameters of nuclear matter and the parameter space of the NJL model,
	we construct two datasets of thermodynamically consistent and
	causal hybrid EoSs, compatible with astrophysical observations.
	We conclude that, to sustain a considerable quark core size, 
	the intensity of the phase transition from hadron to quark matter cannot be strong, having a energy density gap below $200$  MeV/fm$^3$,
	and must occur at baryon densities not above four times the saturation density. A non zero but not too strong quark vector-isoscalar  term and a weak  vector isovector quark term are required. Large quark cores carrying almost half of the star mass are possible inside neutron stars with a maximum mass $\approx 2.2 M_\odot$.
	To get a considerable number of hybrid EoS predicting quark matter already inside neutron stars with a mass $\sim 1.4 M_\odot$, we
	require that the onset of quarks occurs in the range $1.3n_0$ and $2.5n_0$.  Neutron stars with large quark cores corresponding to more than one fourth of the total star mass, are possible if the energy density gap and the pressure at transition are below 100 MeV/fm$^3$.  However, under these constraints, the maximum neutron star mass  is limited to $\lesssim 2.06 M_{\odot}$.  
	No strong signatures from quark matter were found on the radius and the tidal deformability for neutron star masses below $1.8\, M_\odot$.

\end{abstract}

\maketitle

\section{Introduction}
\label{introduction}

Since its discovery in 1967 \cite{Hewish1968}, neutron stars (NS) have been the focus of many experimental and theoretical studies in astrophysics, nuclear and particle physics, due to its extreme properties not attainable in terrestrial laboratories. However, the inner composition of these objects remains unknown and the neutron star equation of state (EoS), which encodes these objects composition, is only known for low densities i.e., for the crust of the star. 
As a result of the extreme densities reached in the core of these objects,  exotic matter including hyperons, Bose-Einstein condensates or quark matter, may exist \cite{glendenning2012compact}. In the latter scenario, baryons begin to overlap and baryonic matter might undergo a phase transition to quark matter \cite{Glendenning1992}.

In recent years multi-messenger astrophysics have been providing a deeper insight on some properties of these objects by combining information from astrophysical observations and, more recently, gravitational wave (GW) information coming from the binary neutron star merger GW170817, provided by the LIGO/Virgo collaborations \cite{TheLIGOScientific:2017qsa,PhysRevLett.121.161101}. Another source of observational data is coming from the Neutron Star Interior Composition Explorer (NICER) experiment by NASA.
NICER was able to estimate the mass and radius of the  millisecond-pulsar PSR J0030+0451 and also infer some thermal properties of hot regions present in the star \cite{Riley_2019}.

The two solar mass pulsars PSR J1614-2230 ($M=1.908\pm$0.016 $M_\odot$) and PSR J0348+0432 ($M=2.01\pm$0.04 $M_\odot$) \cite{Antoniadis:2013pzd} allied to the gravitational wave signal  coming from the neutron star binary merger GW100817 event, define substantial constrains on the EoS, both on the maximum mass and the tidal deformability of the star.  The GW detection was supplemented by the follow up of the electromagnetic counterpart, the gamma-ray burst (GRB) GRB170817A \cite{grb}, and the
electromagnetic transient AT2017gfo \cite{kilo}, that set extra
constraints on the lower limit of the tidal deformability
\cite{Radice2017,Radice2018,Bauswein2019,Coughlin2018,Wang2018}. 
The denser region of the EoS is severely constrained, being very difficult to build models with exotic degrees of freedom inside the stars. More precisely, soft EoS at high densities were ruled out by the discovery of the above-mentioned massive pulsars while EoS that are too stiff and have large radii are incompatible with the tidal deformability coming from  GW observations \cite{Alford:2019oge}. One way to balance both these features might be with an EoS describing a first order phase transition, which would be soft enough at low densities, satisfying the constrains from GW physics and stiff enough after the transition to attain stars with a sufficiently high mass \cite{Alford:2019oge}.

The major difficult on inferring the presence of exotic matter inside neutrons stars, lays on detecting observational signatures that clearly separate a neutron star described by a purely hadronic EoS from an EoS with exotic degrees of freedom. In the case of hadron and quark matter, even though there is a clear physical distinction between the two, in practice, it is very difficult to distinguish the effects of each  type of matter using observables as the star mass, radius and tidal deformability. As discussed in \cite{Most:2018eaw,Alford:2019oge,Weih:2019xvw}, the presence of a first order phase transition between hadronic matter and quark matter can lead to observational signatures that could be exploited in more neutron star binary mergers or observations, favoring the hypothesis of quark matter in the neutron star core.

In order to study matter under such extreme conditions and the possibility of the existence of neutron stars with a quark core, theoretical models must take into consideration that the low density equation of state (EoS), near the saturation, is dominated by hadron degrees of freedom while, at high densities, quarks are the relevant degree of freedom. Hence, to study the possibility of neutron stars with a quark core, hybrid models should be employed to build the hybrid star EoS.

The difficulty of calculating the neutron star EoS from first principles, using the theory of strong interactions, Quantum Chromodynamics (QCD), is due to the non-perturbative properties of such theory for large baryon densities and small temperatures, the physical conditions present inside the NS. The application of Monte Carlo methods on a lattice to QCD can circumvent the non-perturbative nature of the theory but, currently, are not able to perform simulations at high finite density and low temperatures due to the so-called sign problem \cite{Schmidt:2017bjt}. 

One way to study the NS EoS, is to build effective models which incorporate the most important features of the strong interactions within certain limits of applicability. In this work, to study the possibility of formation of a quark core and the signatures of its existence, a two-model approach to the description of the EoS of compact stars is undertaken, i.e., a hybrid equation of state with a hadron phase connected to a quark phase through a first-order phase transition, based in different models for each phase. This two-model approach has been widely used in previous works where an hadronic model and an independent quark model were considered to build a neutron star EoS, see \cite{Pagliara:2007ph,Benic:2014iaa,Benic:2014jia,Zacchi:2015oma,Pereira:2016dfg,Wu:2018kww}.

The description of the hadron phase is well known around the saturation density, but at high densities our knowledge about the EoS is very limited. Several techniques have been applied throughout the last decades to describe this type of matter such as: Skyrme interactions, relativistic mean field models and Taylor expansions around saturation density \cite{Brink1971,Douchin2001,Ducoin2011,Dutra2012,Dutra2014,Margueron2018a,Zhang2018,Malik2019,Li:2019sxd}. One  feature that these approaches have in common is  that free parameters are usually fixed to reproduce nuclear properties at saturation. In this work we will consider hadronic EoSs built using the Taylor expansion. The space of possible parameters will be sampled and several meta-models will be considered, as long as they fulfill a given number of well established conditions.
 The reason is simple: there is still uncertainty in the hadronic phase and using a meta-model approach which does not rely on any specific physical model will allow us to study the possibility of the existence of a quark matter core in a hadron-model independent way.

For the quark phase, the MIT bag model or Nambu$-$Jona-Lasinio (NJL) type models have been widely used to study hybrid neutron stars \cite{Chodos:1974pn,Schertler:1999xn,Baldo:2002ju,Shovkovy:2003ce,Pagliara:2007ph,Bonanno:2011ch}. In the case of the NJL model it has been verified that the appearance of stable quark matter inside the star is very dependent on the model parameters.  In \cite{Buballa:2003et} it was argued that the hadron-quark phase transition is controlled by the light quarks effective mass in the vacuum: smaller values shifts the zero pressure towards lower chemical potentials, favouring the appearance of stable quark matter for massive stars. One can also define an effective bag constant in the NJL model which has a similar effect. It has also been discussed that vector interaction channel, connected to the excitations of vector and pseudovector mesons, stiffens the quark EoS and its effect has been widely studied in the context of neutron stars (see for example \cite{Hanauske:2001nc,Klahn:2006iw,Pagliara:2007ph,Bonanno:2011ch,Lenzi:2012xz,Masuda:2012ed,Klahn:2013kga,Logoteta:2013ipa,Menezes:2014aka,Klahn:2015mfa,Pereira:2016dfg}). Building a stiff quark EoS is also important to describe a quark core inside the star since, at high densities, a stiff quark EoS will be able to sustain stars with a larger maximum mass. The coupling constants of the vector channels can be fixed to vector meson masses in the vacuum \cite{Klimt:1989pm,Lutz:1992dv}, however in-medium effects might change the overall magnitude of these interactions \cite{Fukushima:2008wg}.

In the present study we describe quark matter within the SU(3) NJL model, with the Kobayashi-Maskawa-’t Hooft interaction, vector-isoscalar, and vector-isovector interactions. The employed parametrization yields a small effective light quark mass \cite{Pereira:2016dfg}. We will also consider different values for both the vector-isoscalar and vector-isovector interactions and an effective bag constant.

In this work, we will build hybrid models based on the following assumptions: causality, thermodynamical consistency, including a first order phase transition from hadron matter to quark matter, a hadron phase described by models with good empirical properties at the saturation density,  chiral symmetry in the quark phase, beta equilibrium and global charge neutrality. The ultimate goal is to study possible observational signatures of the presence of a quark core inside a neutron star. To this end we aim to build hybrid EoSs where the hadron part is model independent while the quark phase is descried by a relativistic model based on the essential properties of quark matter. We also explore the connection between stable, massive hybrid stars and the overall magnitude of the phase transition between hadron and quark matter.

This paper is organized as follows: in Section \ref{model_and_formalism} the details of the hadron and quark phases are discussed and the approach to build hybrid equations of state is laid out. In Section \ref{results} the results are discussed and in Section \ref{conclusions} we conclude and some further work is planned.

\section{Model and Formalism}
\label{model_and_formalism}

In order to study hybrid stars, the neutron star equation of state must be described by two distinct phases: an hadronic phase at low densities and a quark phase in the high density region.  

Since our goal is to build hybrid equations of state to study the possible existence of a quark core inside neutron stars and its consequences for the star properties, the hybrid model should be as independent as possible from the choice of the hadronic part of the EoS, at low densities. In order to accomplish this, we consider for the hadronic phase a fourth-order Taylor expansion of the energy functional around the saturation density. Within our approach, there will be eight free coefficients in the expansion that can be related to nuclear properties at saturation. The Taylor expansion approach has some drawbacks: it may lead to non-relativistic equations of state where the in-medium speed of sound is larger than the speed of light or to models which are not thermodynamically stable e.g., the EoS is not a monotonically increasing function of density. In order to tackle these problems, only hadron EoS with good physical properties up to the hadron-quark transition density will be considered as valid and will be used to build hybrid EoS.

For the quark phase,  the SU(3) NJL model is considered. This model is widely used as an effective model of QCD at finite temperature and density, having the same global symmetries as QCD. It reproduces the spontaneous breaking of chiral symmetry in the vacuum and its restoration at some high temperature and/or high density. The main disadvantage of this model is the lack of deconfinement physics: the mesons are bound states of quark-antiquark pairs which can decay at some high energies. Using the three flavor version of the model will also allow to study the possible appearance of strangeness degrees of freedom inside the star. Both hadron and quark phases are considered with zero global electric charge and in beta equilibrium. We neglect the contribution coming from neutrinos since we can consider that they escaped the neutron star during the cooling process.

Even if there is a phase transition from hadron to quark degrees of freedom at some density, it could happen through a first-order phase transition, a second-order phase transition, or via a crossover. The description of first-order phase transition depends on the surface tension between the nuclear-quark phases \cite{Christiansen1997,Maruyama2007,Yasutake2014}, which is still uncertain. Two extreme cases are often considered: the Gibbs construction for low surface tensions and the Maxwell construction for high surface tensions \cite{glendenning2012compact}. In the present work we use only the Maxwell construction.

To build the hybrid EoS using the Maxwell construction, we apply the Gibbs conditions to calculate the transition chemical potential: both phases must be in chemical, thermal and mechanical equilibrium e.g.,
\begin{align}
\mu_B^H & = \mu_B^Q   ,
\\ 
p_B^H &= p_B^Q ,
\\
T_B^H &= T_B^Q,
\end{align}
where the $H$ and $Q$ represent, the hadron and quark phases.

\subsection{Hadronic phase}

The energy per particle functional of homogeneous nuclear matter, $\mathcal{E}$, as a function of the neutron and proton densities, $n_n$ and $n_p$, can be written as:
\begin{align}
\mathcal{E} \qty(n_n,n_p) = 
e_{sat}\qty(n) +
e_{sym}(n)\delta^2 .
\end{align}
where $n=n_n+n_p$ is the baryonic density and $\delta=(n_n-n_p)/n$ is the asymmetry. 
Following previous studies \cite{Margueron2018a,Margueron2018b, Zhang:2018vrx, Zhang:2018bwq, Zhang:2018vbw,Margueron2019}, we parametrize
the hadronic EoS as a Taylor expansion.
 We consider a fourth order expansion around saturation density, $n_{0}$:
\begin{align*}
e_{sat}(x) & = 
E_{sat} + 
\frac{1}{2} K_{sat}x^2 + 
\frac{1}{6} Q_{sat}x^3 + 
\frac{1}{24}Z_{sat}x^4 ,
\numberthis
\\
e_{sym}(x) & = 
E_{sym} + 
L_{sym}x + 
\frac{1}{2}K_{sym}x^2 + 
\frac{1}{6}Q_{sym}x^3\\
& + 
\frac{1}{24}Z_{sym}x^4 .
\numberthis
\end{align*}
Here, $x=(n-n_{0})/(3n_{0})$.  The coefficients of $e_{sat}(n)$ can be identified with isoscalar empirical parameters:
\begin{align}
P_{IS}^{(k)} = 
(3n_{0})^k\left.\frac{\partial^k e_{sat}}{\partial n^k}\right|_{\{\delta=0,n=n_{0}\}} .
\end{align}
While the coefficients of $e_{sym}(n)$ are related to isovector parameters through
\begin{align}
 P_{IV}^{(k)} = (3n_{0})^k\left.\frac{\partial^k e_{sym}}{\partial n^k}\right|_{\{\delta=0,n=n_{0}\}}.
\end{align}

The correspondence between the coefficients and the empirical parameters can then be written as:
\begin{align}
&\qty{E_{sat},K_{sat},  Q_{sat},Z_{sat} } \nonumber \\ 
& \rightarrow \qty{P_{IS}^{(0)},P_{IS}^{(2)},P_{IS}^{(3)},P_{IS}^{(4)} } ,\\
&\qty{E_{sym},L_{sym}, K_{sym},  Q_{sym},Z_{sym} } \nonumber\\
& \rightarrow \qty{P_{IV}^{(0)},P_{IV}^{(1)},P_{IV}^{(2)},P_{IV}^{(3)},P_{IV}^{(4)}}.
\end{align}

The higher order coefficients, $Q_{sat},\, Z_{sat}$ and $K_{sym}, \, Q_{sym},  \, Z_{sym}$ are still poorly known \cite{Farine1997,De2015,Mondal2016,Margueron2018b,Malik2018,Zhang:2018vrx,Li2019}, while the low order ones are better constrained by experimental results. The saturation energy density $E_{sat}$ and saturation density $n_{0}$ are also well known from experiments and we fix their values to: $E_{sat}=-15.8$ MeV (the current estimated value is  $-15.8\pm0.3$ MeV \cite{Margueron2018a}), and $n_{0}=0.155$ fm$^{-3}$.

To build an hadronic EoS, we use random sampling to choose a point in the 8-dimensional space of parameters from a multivariate Gaussian with zero covariance:
\begin{align}
\text{EoS}_i &= (E_{sym},L_{sym},K_{sat},K_{sym},Q_{sat},Q_{sym},Z_{sat},Z_{sym})_i\nonumber\\
&\sim N(\boldsymbol{\mu},\boldsymbol{\Sigma})
\label{eq:hadronic1}
\end{align}
where 
\begin{equation}
\boldsymbol{\mu}^T=(\overline{E}_{sym},\overline{L}_{sym},\overline{K}_{sat},\overline{K}_{sym},\overline{Q}_{sat},\overline{Q}_{sym},\overline{Z}_{sat},\overline{Z}_{sym}),
\label{eq:hadronic2}
\end{equation}
is the mean vector, and
\begin{align}
\boldsymbol{\Sigma}=diag(\sigma_{E_{sym}},&\sigma_{L_{sym}},\sigma_{K_{sat}},\sigma_{K_{sym}},\nonumber\\
&\sigma_{Q_{sat}}, \sigma_{Q_{sym}},\sigma_{Z_{sat}},\sigma_{Z_{sym}}).
\label{eq:hadronic3}
\end{align}
is the covariance matrix. The values used are in Table \ref{table_gaussian_dist}.

\begin{table}[ht]
\begin{center}
\begin{tabular}{lllllllll}
\hline
$P_{i}$  & $E_{sym}$ &  $L_{sym}$ & $K_{sat}$ & $K_{sym}$ & $Q_{sat}$ & $Q_{sym}$ & $Z_{sat}$ & $Z_{sym}$ \\
 \hline
  \hline
$\overline{P}_{i}$ & $32$  & $60$ & $230$ & $-100$ & $300$ & $0$& $-500$ & $-500$ \\
$\sqrt{\sigma_{{P}_{i}}}$  & $2$ & $15$ & $20$ & $100$ & $400$ & $400$& $1000$ & $1000$\\
\hline
\end{tabular}
\end{center}
\caption{The mean $\overline{P}_{i}$ and standard deviation
$\sqrt{\sigma_{{P}_{i}}}$ of the multivariate Gaussian, where
$\sigma_{{P}_{i}}$ is the variance of the parameter $P_{i}$. 
Our EoSs are sampled using the initial distribution for 
$P_i$ assuming that there are no correlations
among the parameters. All the quantities are in units of MeV.
The values of $E_{sat}$ and $n_{0}$ are fixed to $-15.8$ MeV and $0.155$ fm$^{-3}$,
respectively.}
\label{table_gaussian_dist}
\end{table}

Using this method, no correlations exist between the coefficients of the Taylor expansion and, consequentially, between the empirical parameters \cite{Margueron2018a,Margueron2018b,Zhang:2018vrx,Ferreira:2019bgy}. Building the hybrid EoS and applying to it physical, experimental and observational constraints will give rise to correlations between the empirical parameters. There will also be correlations between the empirical parameters of the hadronic model and the ones used to build the quark models

\subsection{Quark phase}

In the NJL Lagrangian, we will consider only chiral symmetry conserving interactions: the usual scalar and pseudo-scalar four-Fermi interaction, the 't Hooft determinant (which breaks the $U_A(1)$ symmetry) and chirally symmetric vector interactions. Regarding the vector interactions, previous works have showed their importance in the stiffening of the EoS and in building models that can reach two-solar mass stars \cite{Pereira:2016dfg}. As in \cite{Pereira:2016dfg}, we will consider the vector and pseudovector interaction (vector-isoscalar) and the vector-isovector and
pseudovector-isovector interaction (vector-isovector) with distinct couplings, $G_\omega$ and $G_\rho$, respectively. These interactions have the same quantum number as the $\omega$ and $\rho$ mesons and the couplings can be fixed to their masses \cite{Klimt:1989pm}. However these quark bilinears are proportional to density and in-medium effects can change the overall magnitude of these interactions. Hence, as in \cite{Pereira:2016dfg}, each ratio $\xi_\omega = G_\omega / G_S$ and $\xi_\rho = G_\rho / G_S$ will define a different model, with different properties. The Lagrangian density can be written as:
\begin{align*}
\mathcal{L} & = 
\bar{\psi} 
\qty(
i\slashed{\partial} - \hat{m} + \hat{\mu} \gamma^0 
) 
\psi 
\\
& + G_S  \sum_{a=0}^8
\qty[ \qty(\bar{\psi} \lambda^a \psi)^2 + 
\qty(\bar{\psi} i \gamma^5 \lambda^a \psi)^2 ]
\\
& - G_D \qty[  
\det\qty( \bar{\psi} \qty(1+\gamma_5) \psi ) + 
\det\qty( \bar{\psi} \qty(1-\gamma_5) \psi )  ]
\\
& - G_\omega\Big[ (\bar{\psi}\gamma^\mu\lambda^0\psi)^2 + (\bar{\psi}\gamma^\mu\gamma_5\lambda^0\psi)^2 \Big]
\\
& - G_\rho \sum_{a=1}^8 
\qty[ 
(\bar{\psi} \gamma^\mu\lambda^a \psi)^2 +  
(\bar{\psi} \gamma^\mu\gamma_5\lambda^a \psi)^2 
]
.
\numberthis
\label{eq:SU3_NJL_lagrangian}
\end{align*} 
Where $\psi$ is a $N_f$-component vector in flavor space, $\hat{m}=diag\qty(m_u, m_d, m_s )$ and $\hat{\mu}=diag\qty(\mu_u, \mu_d, \mu_s )$ are the quark current mass and chemical potential matrices and $\lambda^a$ ($a=1,2...8$) are the Gell-Mann matrices of the SU(3) group and $\lambda^0=\sqrt{\nicefrac{2}{3}} \mathds{1}$. The determinant is to be carried out in flavor space. The NJL model is not re-normalizable meaning that some regularization procedure must be employed. In this work we use the 3-momentum cutoff scheme. The thermodynamics of the model is obtained
in the mean-field approximation (see appendix for details).

We use the vacuum parameter set of Table \ref{tab:2} that was proposed in \cite{Pereira:2016dfg}, which provides an effective mass for the light quarks of approximately one third ot the nucleon mass, $M_u=M_d\approx 313$ MeV. In Table \ref{tab:3}, we compare the values of the calculated observables with the respective experimental values.

\begin{table}[ht!]
\begin{ruledtabular}
\begin{tabular}{cccccccc}
$\Lambda$  & $m_{u,d}$ & $m_s$  & $G_S\Lambda^2 $ & $G_D\Lambda^5 $ & $M_{u,d}$ & $M_s$ \\
\text{[MeV]}      &  [MeV]    & [MeV]   &               &                  & [MeV] &  [MeV]\\
\hline
630.0 & 5.5   & 135.7 & 1.781 &  9.29 & 312.2   & 508 \\
\end{tabular}
\end{ruledtabular}
\caption{$\Lambda$ is the model cutoff, $m_{u,d}$ and $m_{s}$ are the quark current masses, $G_S$ and $G_D$ are coupling constants. $M_{u,d}$ and $M_{s}$ are the resulting constituent quark masses in the vacuum.}
\label{tab:2}
\end{table}

\begin{table}[ht!]
\begin{ruledtabular}
    \begin{tabular}{ccc}
     & NJL SU(3) {   }& Experimental \cite{Agashe:2014kda} \\
        \hline
    $m_{\pi^{\pm}}$ [MeV]     & 138.5  & 139.6 \\
    $f_{\pi^{\pm}}$ [MeV]     & 90.7   & 92.2 \\
    $m_{K^{\pm}}$ [MeV]       & 493.5  & 493.7 \\
    $f_{K^{\pm}}$ [MeV]       & 96.3   & 110.4 \\
    $m_{\eta}$ [MeV]    & 479.1  & 547.9 \\
    $m_{\eta'}$ [MeV]   & 837.9  & 957.8 \\
    \end{tabular}
\end{ruledtabular}    
    \caption{Masses and decay constants of several mesons within the model and the respective experimental values.}
  \label{tab:3}
\end{table}

As remarked in \cite{Pagliara:2007ph}, the NJL model pressure and energy density are defined up to a constant $B$, analogous to the MIT bag constant. When using the NJL model, this constant has been shown to be essential to build hybrid EoS which sustain two-solar mass neutrons stars. In \cite{Pagliara:2007ph,Pereira:2016dfg}, the bag constant was fixed by requiring that the deconfinement occurs at the same baryonic chemical potential as the chiral phase transition. More recently in \cite{Han:2019bub}, an effective bag constant was also used to control the density at which the phase transition from hadron to quark matter happened. Similar to the last approach, different values for the bag constant will be considered and the general effect of increasing $B$ will be studied. Different values of the bag constant will mean different onsets of phase transition for the same combination of hadron and quark models. To include the effect of a finite bag constant, the quark EoS is modified: $P \to P + B$ and $\epsilon \to \epsilon-B$.
Hence the NJL quark EoS will be defined by three parameters: the model vector coupling ratios, $\xi_\omega= G_\omega / G_S$, $\xi_\rho=G_\rho / G_S$ and the bag constant $B$.

\subsection{The Hybrid EoS}

In this work we will consider a set of distinct hybrid equations of state spanned by different hadron and quark models. 
To build one particular hybrid EoS, we start with the hadronic part. First, a point in the 8-dimensional multivariate Gaussian 
distribution is randomly chosen, using  Eqs. (\ref{eq:hadronic1})-(\ref{eq:hadronic3}), defining an hadronic EoS. 
Valid hadronic EoS must cross the Sly4 EoS \cite{Douchin2001} in the
$P(\mu)$ plane below $n<0.10$ fm$^{-3}$, consistently with the range of core-crust transition densities for a large set of nuclear models \cite{Ducoin2011}.
The third step is looking for possible quark models, spanned by the parameters $\qty( \xi_\omega, \xi_\rho, B )$, that 
that allow for a first order phase transition, through a Maxwell construction, to quark matter.
Note that multiple hybrid EoS might be described by the same hadronic part but with different quark matter phases.

A valid hybrid EoS must satisfy the following conditions:
i) $p(\epsilon)$ is  monotonically increasing (thermodynamic stability);
ii) the speed of sound must not exceed the speed of light (causality);
iii) a maximum mass at least as high as $1.97M_{\odot}$ must be supported (observational constraint \cite{Arzoumanian2017,Fonseca2017,Demorest2010,Antoniadis:2013pzd});
iv) the hadron-quark phase transition density  {\bf of $\beta$-equilibrated matter} must occur at $n>0.2$ fm$^{-3}$  (quark matter has not been detected close to the saturation density or below).
If a given hybrid EoS fails one of these tests it is rejected. If it passes these tests, the mass-radius relations and tidal deformability for static and spherically symmetric stars are calculated using the Tolman–Oppenheimer–Volkoff equations and tidal deformability equation \cite{glendenning2012compact,Hinderer:2007mb,Postnikov:2010yn}. 

This process is then repeated for  different hadronic models to build a dataset containing several hybrid EoS parametrized by 11 parameters: 8 coming from the hadronic phase and 3 from the quark phase.

\section{Results}
\label{results}

Following the formalism presented in the previous section, we have searched for possible first-order phase transitions, through a Maxwell construction, from hadronic to quark matter. The quark matter is parametrized by $\qty( \xi_\omega, \xi_\rho, B )$ while the hadronic part is characterized 
by the following empirical parameters of nuclear matter $(E_{sym},L_{sym},K_{sat},K_{sym},Q_{sat},Q_{sym},Z_{sat},Z_{sym})$.

Two distinct datasets will be constructed within the following sections.
Firstly, we will look for hybrid EoS where a first-order phase transition from hadron to quark matter
happens at any density above $0.2$ fm$^{-3}$. In a second step, we will also set an upper density bond, i.e., we define
a density region where the transition must occur.  

\subsection{Hybrid EoS set with phase transitions for $n>0.2$ fm$^{-3}$}

We have generated 23723 hybrid EoS where a first-order phase transition to quark matter occurs at densities $n>0.2$ fm$^{-3}$. 
The quark models were searched in a fixed grid:
for the vector couplings we consider $\xi_\omega \in [0,1]$ with $0.1$ intervals, while for the $\xi_\rho$ coupling we consider the same range but with $0.2$ intervals. For the bag constant we consider four different possible values: $B=\qty{0,5,10,15}$ MeV fm$^{-3}$.

\subsubsection{$M$-$R$ and $M$-$\Lambda$ diagrams}

\begin{figure}[!ht]
	\centering
	\includegraphics[draft=false,width=1.0\columnwidth]{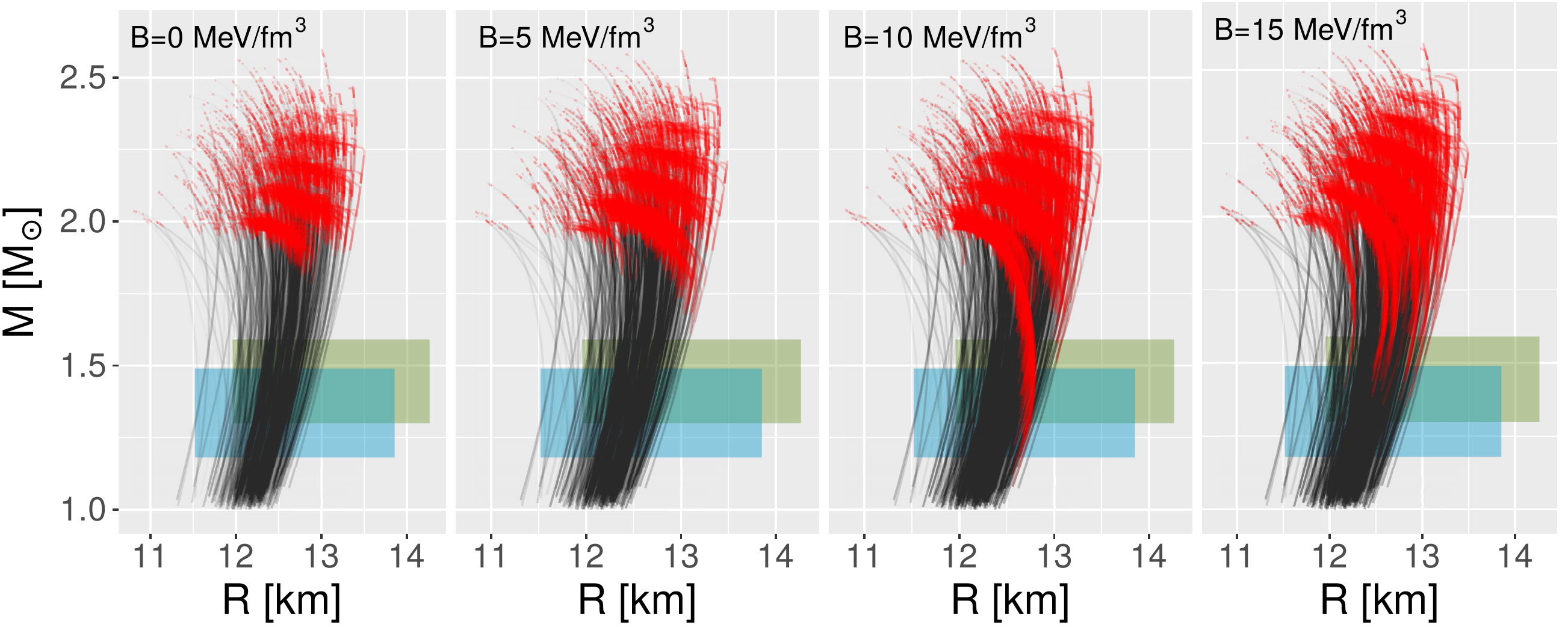}
	\caption{Mass-radius diagrams for the entire dataset. Each panel has a specific value for the bag constant, $B= 0,5,10,15$ MeV fm$^{-3}$. 
		The black lines represent the hadronic branches while the red indicates the quark branches of stable stars.
	The colored regions indicate the $(M,R)$ constraints obtained by two independent analysis using 
	the NICER x-ray data from the millisecond pulsar PSR J0030+0451 \cite{Riley_2019,Miller:2019cac}.}
	\label{fig:TOV}
\end{figure}

We show the $M(R)$ diagram for our entire dataset in Fig. \ref{fig:TOV}. 
The black lines represent the sequence of purely hadronic NS, whilst the red lines
show the stable hybrid NS branches, i.e., NS that have a quark core.
The presence of a first-order phase transition imprints characteristic features in the mass–radius curve. The sharp transition from hadron matter to quark matter leads to a branch of stable hybrid stars. 
For convenience, we present in four different panels the results for each value of the bag constant, $B=\qty{0,5,10,15}$ MeV fm$^{-3}$, of the quark phase. 
In the present work, only stable connected branches of hybrid stars were found, i.e., the branch of stable hybrid stars is connected to the hadronic branch. 
Disconnected hybrid branch, or ``third family'' of stars are different possibilities that have been studied in other works \cite{Alford:2013aca,Benic:2014jia,Alford:2017qgh,Paschalidis:2017qmb,Han:2018mtj,Montana:2018bkb,Li:2019fqe}.\\

The uncertainty on the hadronic EoS enables the existence of a core of quarks for a wide range of NJL parametrizations.
However, the majority of the hybrid EoS give rise to  very small quark
branches (red lines), meaning that only very small quark cores are allowed before they become unstable.
The length of the quark branch can be measured by $\Delta M = M_{max}-M_t$,
where the $(M_t,R_t)$ is the NS in the $M(R)$ diagram for which the onset of quarks takes place, 
i.e., the lightest NS with non-zero quark content. The mean value for $\Delta M$ of the whole set is $0.03M_{\odot}$ while the minimum/maximum are
$10^{-4}M_{\odot}$/$0.79M_{\odot}$. Thus, the distribution is highly concentrated 
on very small values for the quark branches sizes, in accordance with the value $\lesssim 10^{-3}M_{\odot}$ in \cite{Alford:2013aca,Alford:2015gna,Ranea-Sandoval:2015ldr,Han:2019bub}.  
The length of the quark branch is highly dependent on the properties of the hadron-quark phase transition \cite{Alford:2013aca,Alford:2015gna}.
The detection of such hybrid branches is highly limited by their small extensions.

The distribution of $M_t$, over the 23723 hybrid EoS, is highly concentrated around its mean value $\bar{M_t}\approx2.14M_{\odot}$ (corresponding to small quark cores, as we will see). 
Even massive NS might have a quark core, such as the recently detected MSP J0740$+$6620 
with a mass $2.14^{+0.10}_{-0.09}M_\odot$ \cite{Cromartie2019}.
From the whole dataset of hybrid EoS, only 29 EoS predict quark matter in a $1.4M_{\odot}$ NS, all with a bag constant of $B=10$ MeV/fm$^{3}$ (they are noticeable in Fig. \ref{fig:TOV}). The radius of  all NS obtained,  both light and massive NS, i.e.,  $M>1M_{\odot}$, ranges between $10.7$ km and $13.5$ km.

One conclusion upon analyzing this figure is the effect of the bag constant: as in previous studies \cite{Pereira:2016dfg}, the existence of a bag constant and its overall magnitude is essential in building models which predict larger quark phases inside the star. Increasing the bag constant  shifts the pressure of the quark phase to higher density values. 
This in turn, lowers the critical baryon density for the onset of the transition, $n_{t}$, by effectively decreasing the chemical potential at which both phases are in chemical equilibrium, $\mu_{H}=\mu_{Q}$. The effect of having stars with larger quark phases, with increasing $B$, is also obvious: the phase transition to quark matter softens the EoS, predicting stars with smaller radii, i.e., more compact stars. 
Actually, some quark matter parametrizations already predict $n_{t}<0.2$ fm$^{-3}$ for $B=15$ MeV fm$^{-3}$ (which we do not consider), explaining why families of small NS masses, which are seen for the $B=10$ MeV fm$^{-3}$ case, are not present for $B=15$ MeV fm$^{-3}$. 
In fig. \ref{fig:TOV}, we have marked two regions corresponding to the $(M,R)$ constraints obtained by two independent analysis using 
the NICER x-ray data from the millisecond pulsar PSR J0030+0451 \cite{Riley_2019,Miller:2019cac}. The set of EoS  generated in the present work are well within both regions.

\begin{figure}[!ht]
	\centering
	\includegraphics[draft=false,width=1.0\columnwidth]{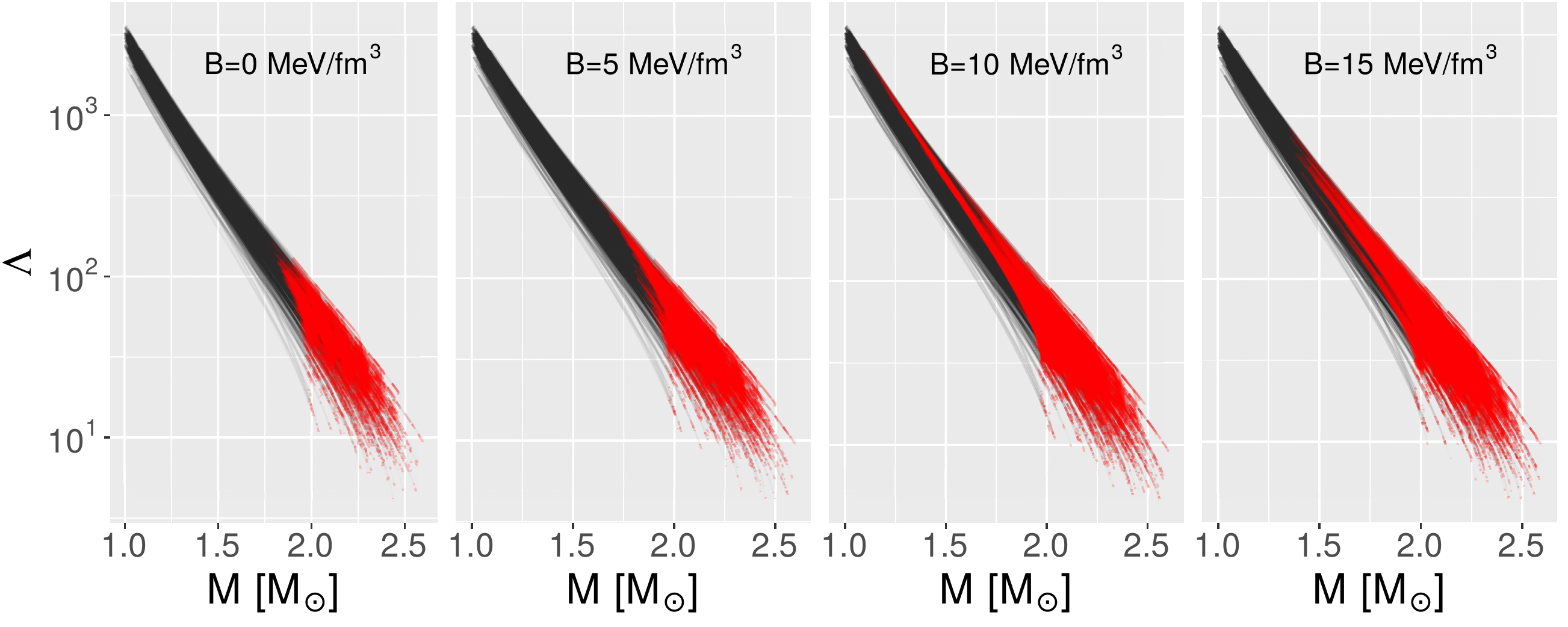}
	\caption{Tidal deformability-mass diagrams for the entire dataset. Each panel has a specific value for the bag constant, $B=0,5,10,15$ MeV fm$^{-3}$. 
		The black lines represent the hadronic branches while the red indicates the quark branches os stable stars.}
	\label{fig:TIDAL}
\end{figure}

The diagrams of the NS tidal deformability as a function of the NS mass are shown in Fig. \ref{fig:TIDAL} (with the same pattern of the last figure). The non-existence of quark matter in light NS, as previously discussed, has a direct consequence in the tidal deformability. Indeed as one can see from Fig. \ref{fig:TIDAL}, the tidal deformability of less massive stars, $M<1.5M_\odot$ is almost completely dictated by the hadron part of the equation of state. 

Being the stiffness of EoS correlated with tidal deformability, the result $70<\Lambda_{1.4M_{\odot}}<580$ (90\% level) from the  GW170817 event \cite{Abbott18} rules out very stiff hadronic EoS. 
On the other hand, having a too soft hadronic EoS is incompatible with observed massive NS. A phase transition from hadronic to quark matter might help in accommodate a stiff hadronic EoS (which is also a requirement for getting a first-order transition), and thus describing massive NS, and at the same time getting low tidal deformabilities for light NSs. The crucial point here is the density at which this phase transition takes place.
If the density is low enough, the softening of the hybrid EoS, originated from the onset of quarks, might reduce the tidal deformability for NS masses around $1.4M_{\odot}$, which otherwise would be too high if only hadronic matter would be present. If the onset of quarks happens at high densities it becomes quite unlikely the presence of quarks in $\approx1.4M_{\odot}$ NS.

The present set of hybrid EoS have a wide distribution for $\Lambda_{1.4M_{\odot}}$, characterized by the mean value of $558$, a standard deviation of $74$, and minimum (maximum) value of $311$ ($724$). Only 14595 hybrid EoS, about 62\%, satisfy $70<\Lambda_{1.4M_{\odot}}<580$. From the whole set of EoS, only 29 hybrid EoS predict quark matter inside a  $1.4M_{\odot}$ NS. Therefore, it is important to 
analyze the properties of the hadronic matter phase, which will be performed in the next section.

\subsubsection{Hadronic matter phase properties}

The set of hybrid EoS we are analyzing was generated from
354 distinct hadronic EoS. Their parameters statistics are given in Table \ref{tab:hadron}.
The results can be compared with \cite{Ferreira:2019bgy}, where a set of purely hadronic EoS  
was constructed using the same formalism and crust (SLy4). 
The comparison is useful in accessing how the presence of a first-order phase transition
to quark matter changes the final distribution of the empirical parameters.   
The two empirical properties that show highest deviations are $Q_{sat}=241.38\pm242.03$MeV and $Z_{sat}=362.11\pm593.75$MeV. These numbers should be compared with
the values $Q_{sat}=56.04\pm122.31$ MeV and $Z_{sat}=-178.46\pm 141.26$ MeV, which were reported in \cite{Ferreira:2019bgy}. 
This indicates that in order to obtain a phase transition to quark matter, the hadronic matter is characterized 
by a higher value $Q_{sat}$ and a higher (and positive) value for $Z_{sat}$, both quantities with a wider spread.

\begin{table}[ht]
	\centering
	\begin{tabular}{ccccccccc}
		\hline
		& $K_{sat}$ & $Q_{sat}$ & $Z_{sat}$ & $E_{sym}$ & $L_{sym}$ & $K_{sym}$ & $Q_{sym}$ & $Z_{sym}$ \\ 
		\hline
mean & 236.52 & 241.38 & 362.11 & 33.17 & 49.99 & -37.48 & 191.54 & 503.39 \\ 
std & 17.73 & 242.03 & 593.75 & 1.84 & 12.55 & 67.23 & 310.48 & 734.69 \\ 
		\hline
	\end{tabular}
	\caption{Mean and standard deviation (Std) of the hadronic EoS set. 
		All the quantities are in units of MeV.}
	\label{tab:hadron}
\end{table}

In Table \ref{fig:utl2}, we summarize the statistics of some properties of a canonical NS with  mass $1.4M_{\odot}$, over the
 354 distinct hadronic EoS used to build the hybrid EoS. From the mean value of central density, $n_{max}$, we understand the difficulty of having quark matter in a NS with mass $1.4M_{\odot}$: within the NJL model, it is highly unlikely to get the onset of quark matter below $2.5n_0$ and, at the same time, have a quark matter EoS  stiff enough to support the existence of a quark core in a stable $1.97 M_{\odot}$ NS. Despite the mean value of $\Lambda_{1.4M_{\odot}}$ being compatible with the upper bound $580$ from LIGO/Virgo collaboration \cite{Abbott18}, 119 hadronic EoS have $\Lambda_{1.4M_{\odot}}>580$. 

At this point we would like to comment that although the hadronic EoS set obtained above does not include explicitly hyperons, we may consider that the effect of these degrees of freedom is accounted for in an effective way. Since the onset of hyperons occurs as a crossover, the meta-models used to describe the hadronic matter may also simulate EoS with hyperons, in the sense that the onset of hyperons softens the EoS. So, the soft EoS   at high densities of our set of hadronic EoS may be due to the presence of hyperons or a soft nucleonic EoS. We consider with the present approach for the hadronic matter we span all possibilities that may originate a quark core.  In particular, in Fig. \ref{fig:TOV} we  may identify some $M(R)$ curves that reproduce the behavior of the $M(R)$ curves expected when hyperons are included, see for instance Fig. 5 of \cite{Fortin16}.  It has been discussed in \cite{Providencia2019,Fortin2020} that  some star properties, as the radius and mass, are affected in a similar way if a soft symmetry energy is considered or hyperonic degrees of freedom are included. These two effects cannot be separated due to the our lack of information on the high density nucleonic EoS.  We may understand the onset of quarks at a density close to twice saturation density, as we will discuss in section Sec. \ref{III3b}, resulting from a possible competition between the onset of quarks of the onset  of hyperons, where the quarks are favored.



\begin{table}[ht]
	\centering
	\begin{tabular}{cccccc}
		\hline
		& $R$ [km] & $\Lambda$ & $k_2$ & $n_{max}/n_0$ & $C$ \\ 
		\hline
mean & 12.40 & 544.66 & 0.10 & 2.50 & 0.17 \\ 
std & 0.25 & 79.43 & 0.01 & 0.22 & 0.00 \\ 
min & 11.56 & 311.07 & 0.08 & 2.08 & 0.16 \\ 
max & 12.93 & 723.93 & 0.12 & 3.32 & 0.18 \\ 
		\hline
	\end{tabular}
	\caption{The mean, standard deviation (Std), and maximum/minimum values for the following properties of a $1.4 M_{\odot}$ NS: radius ($R$), tidal deformability ($\Lambda$),
		love number $k_2$, central density ($n_{max}$), and compactness ($C=GM/c^2R$).
}
	\label{fig:utl2}
\end{table}

Since our hadronic equations of state were built with the assumption of a phase transition to quark matter at high densities, even though the tidal deformability of less massive stars are described by the hadron EoS,  the hadron EoS itself carries information of the fact that, at higher densities, a first-order phase transition to quark matter occurs.

As shown in \cite{Alford:2013aca}, the use of the Gibbs construction moves the onset of quarks to lower densities, while the region of the purely quark mater shifts to higher densities. Thus, contrarily to the Maxwell construction, it shows larger quark content for wider range of NS masses, that in turns decreases their radii and tidal deformabilities. However, as also noted in \cite{Alford:2013aca}, the drastic effect from the abrupt hadron/quark transition, present in the Maxwell construction, is smoothed down and thus quarks degrees of freedom are harder to be distinguishable form purely hadronic ones.

\subsubsection{Quark matter phase properties}

Using the Maxwell construction means having a jump in the thermodynamical quantities as a function of baryonic density. In the following discussion we analyze some results as a function of the onset density of the phase transition, $n_t$ and the overall size of the jump in density, $\Delta n = n_q - n_t$, where $n_q$ denotes the baryonic density at which the pure quark matter sets in. We also inspect some results in light of the critical energy density and pressure for the onset of the transition, $\epsilon_t$ and $p_t$, respectively.

\begin{figure}[!ht]
	\centering
	\includegraphics[draft=false,width=1.0\columnwidth]{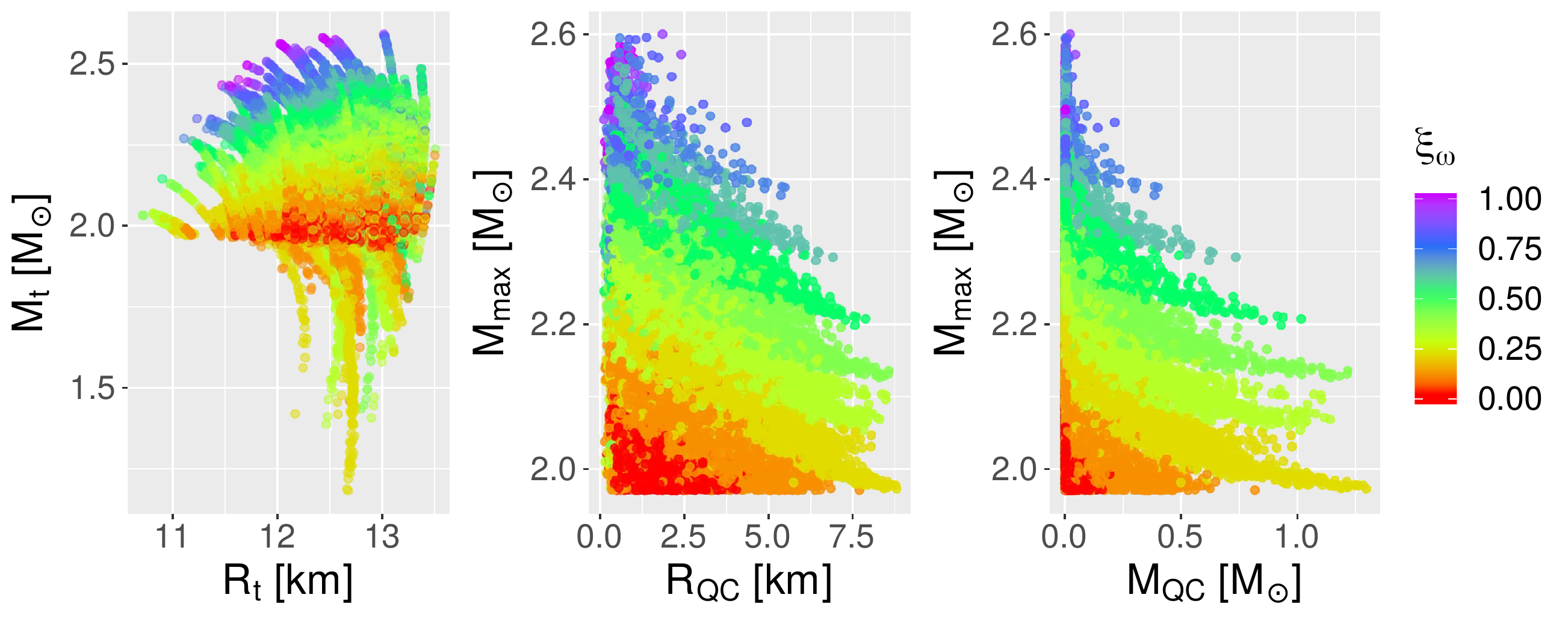}
	\caption{The $M_{t}$ vs $R_{t}$ (left), which corresponds to the $(M,R)$ where the onset of quark matter occurs,
		  and the diagrams maximum NS mass vs. quark core radius (center) and mass (right) as a function of 
		   $\xi_\omega$ (color scale) for each hybrid EoS.}
	\label{fig:result_1}
\end{figure}

Let us now study how the quark matter properties affect the existence of stable hybrid stars. In Fig. \ref{fig:result_1}, we show the effect of the vector-isoscalar parameter $\xi_\omega= G_\omega / G_S$ on the values $(M_t,R_t)$ (left panel) and the radius and mass of the quark core versus its maximum NS mass (center and right panels). Each point represents an hybrid EoS of our set with a specific value $\xi_\omega$, shown in color scale.    
The values $(M_t,R_t)$ give the location of the NS in the $M(R)$ diagram at which quark branches begin  \footnote{the quark branches correspond to the red lines in the Figs. \ref{fig:TOV} and \ref{fig:TIDAL}}, i.e., they characterize the lightest NS inside which quark matter is already present.
Clearly, as $\xi_\omega$ increases, the appearance of quarks occurs for larger NS masses and radii.
Quark cores in massive NS, $M>2.4M_{\odot}$, are attainable with $\xi_\omega>0.6$.
However, when we look to the quark core size and mass (center and right panels), we see that the EoSs that sustain high values of $M_{max}$, 
have a very small quark core. This explains the presence of very small quarks branches for massive stars in the $M(R)$ diagrams of Fig. \ref{fig:TOV}.

\begin{figure}[!ht]
	\centering
	\includegraphics[draft=false,width=1.0\columnwidth]{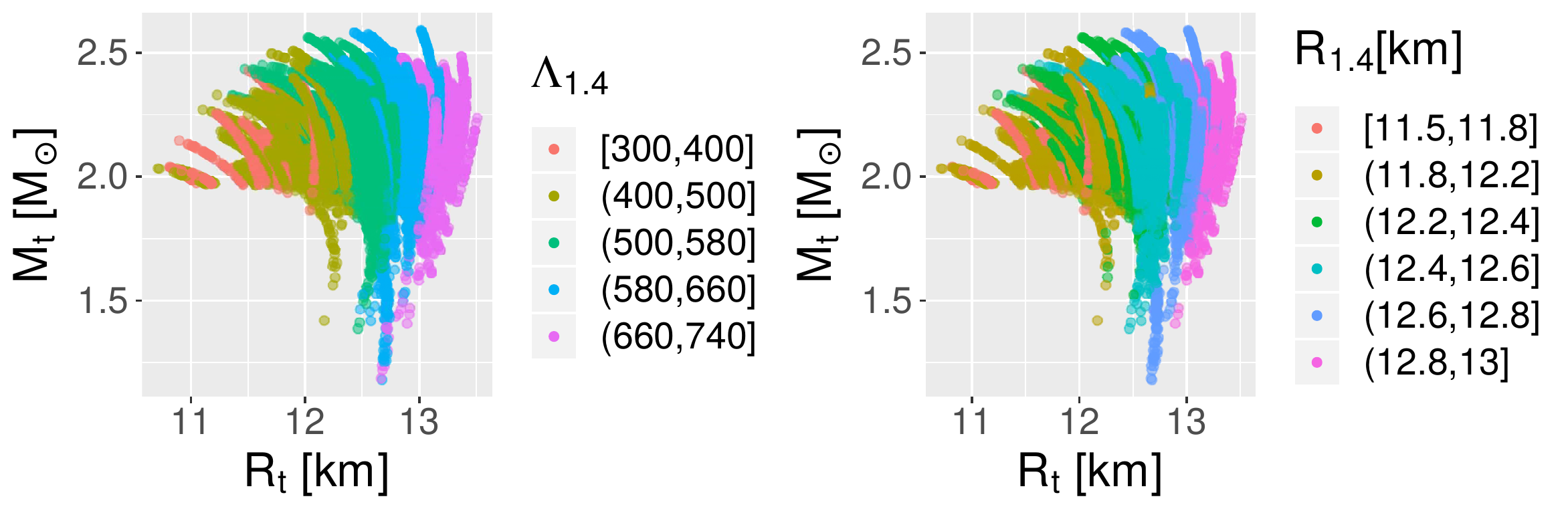}
	\caption{The NS mass and radius corresponding to the onset of quark matter, $M_t$ and $R_t$, for each hybrid EoS.
		The color scale shows the corresponding values of $\Lambda_{1.4M_{\odot}}$ (left) and $R_{1.4M_{\odot}}$ 
		(right) of each hybrid star.}
	\label{fig:parte1_new}
\end{figure}

The radius and mass of the quark core, $R_{QC}$ and $M_{QC}$, is determined from the quark content of the most massive stable NS
that contains quark matter. The distribution of $M_{QC}$ has a mean value of $0.08M_{\odot}$ and a maximum value of 
$1.29M_{\odot}$, whereas the mean value of $R_{QC}$ is $1.99$ km and the maximum value of $8.78$ km. 
Quark cores having radii as large as 7 km have been also obtained in \cite{Annala2019} using a quite different approach to build the EoS and identify the quark phase.
The larger values of  $(R_{QC},M_{QC})$ are generated for $\xi_\omega=0.2$ and $0.3$.

We have already seen that it is highly improbable the existence of quark matter inside light NSs. 
We got only one hybrid EoS with a tiny quark composition for a $1.4M_{\odot}$ NS. 
Being $1.0 \leq M / M_{\odot}\leq 1.8$ the range of interest for Binary Neutron Star (BNS) mergers, the presence of 
small amounts of quarks in stable hybrid makes them almost indistinguishable from a purely hadronic matter scenario.

In Fig. \ref{fig:parte1_new},  two diagrams of $M_t$ versus $R_t$
show,  through color scales, the values $\Lambda_{1.4M_{\odot}}$ (left) and $R_{1.4M_{\odot}}$  (right)  predicted by the EoS corresponding to the pair ($M_t$, $R_t$). Small $\Lambda_{1.4M_{\odot}}$ are associated to smaller $R_t$ values and generally larger $M_t$ because they correspond to softer hadronic EoS, which gives rise to a transition to quark matter at larger densities. We are only able to get $M_t\lesssim1.5 M_{\odot}$ for stiff enough hadronic EoS.

\begin{figure}[!ht]
	\centering
	\includegraphics[draft=false,width=1.0\columnwidth]{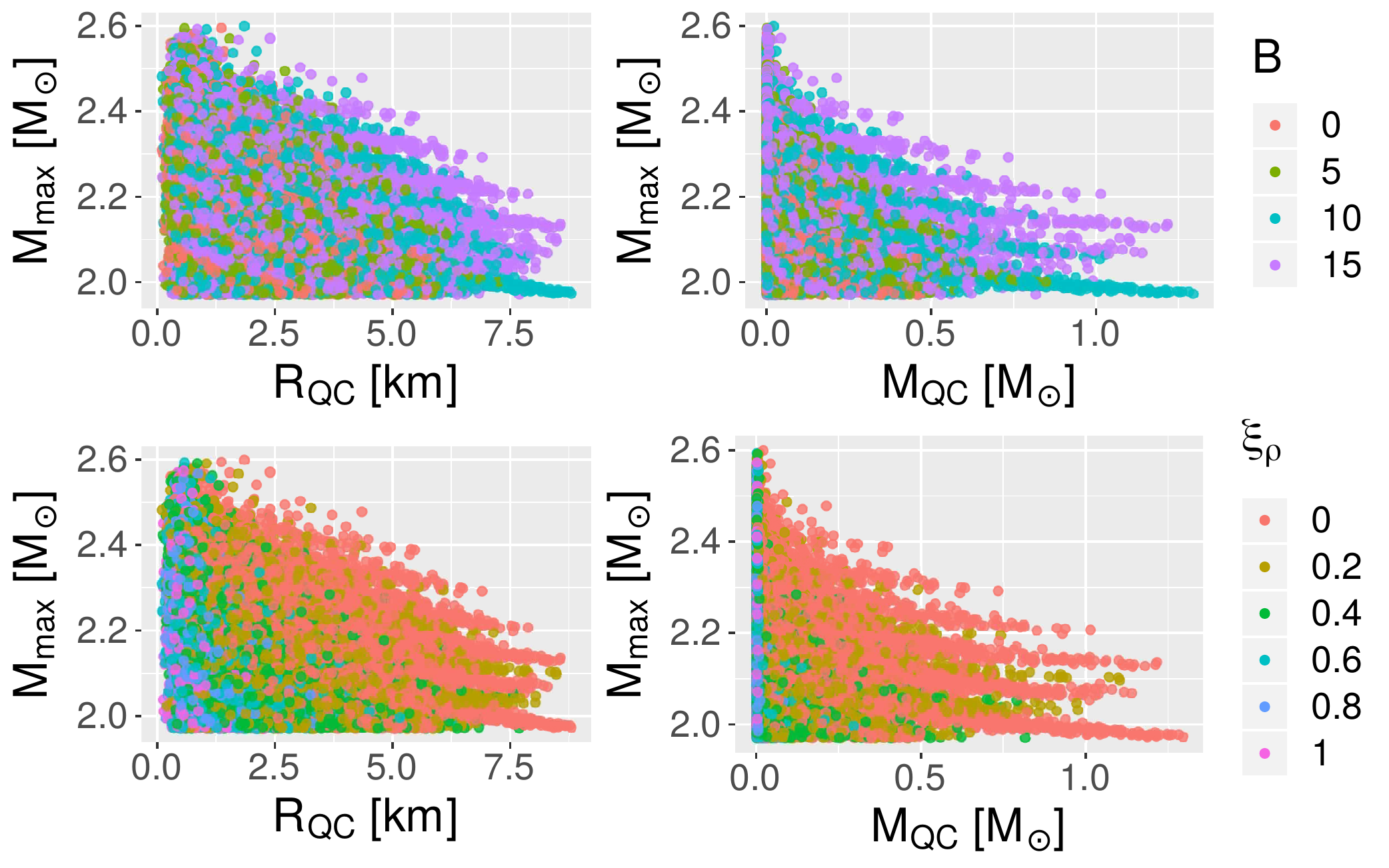}
	\caption{Diagrams maximum NS mass vs. quark core radius (right) and mass (left) as a function of 
	the bag value $B$ [MeV fm$^{-3}$] (top) and $\xi_{\rho}$ (bottom), in color scale, for each hybrid EoS.}
	\label{fig:result_2}
\end{figure}

In the following, we study the general effect of the other two quark matter parameters,
 the bag constant $B$ and the coupling ratio $\xi_\rho= G_\rho / G_S$, on the quark core size
 and the maximum NS mass. The results are in Fig. \ref{fig:result_2}. 
As expected, for a given hadronic EoS, higher values of $B$ correspond to a larger amount of quark matter. 
Hybrid EoS with $B=10$ and $15$ MeV fm$^{-3}$ generate by far the largest cores, i.e., larger values of $R_{QC}$ and $M_{QC}$. 
The effect of $\xi_\rho$ (bottom panels) is also quite clear, larger values give rise to smaller quark cores.
Almost all of the sizable quark cores are obtained for $\xi_\rho<0.4$. 
From both $\xi_\rho$ and $B$ panels, we conclude that, independently of the hadronic EoS properties, 
larger and more massive quark cores are generated for $\xi_\rho<0.4$ and  $B=\{10,15\}$ MeV fm$^{-3}$. 
The vector-isovector interaction, whose strength is controlled by the $\xi_\rho= G_\rho / G_S$ ratio, has the main effect to push the onset of the strange quarks  to lower densities, making the EoS softer at higher densities \cite{Pereira:2016dfg}.
Strangeness makes the EoS softer due to the presence of an extra fermionic  degree of freedom at high densities and, therefore, another Fermi sea to be populated, decreasing the pressure. The consequence of having a softer EoS at high densities means that the star will not be able to sustain large amounts of quark matter in its core, resulting in smaller quark cores for high values of $\xi_\rho= G_\rho / G_S$.

\begin{figure}[!ht]
	\centering
	\includegraphics[draft=false,width=1.0\columnwidth]{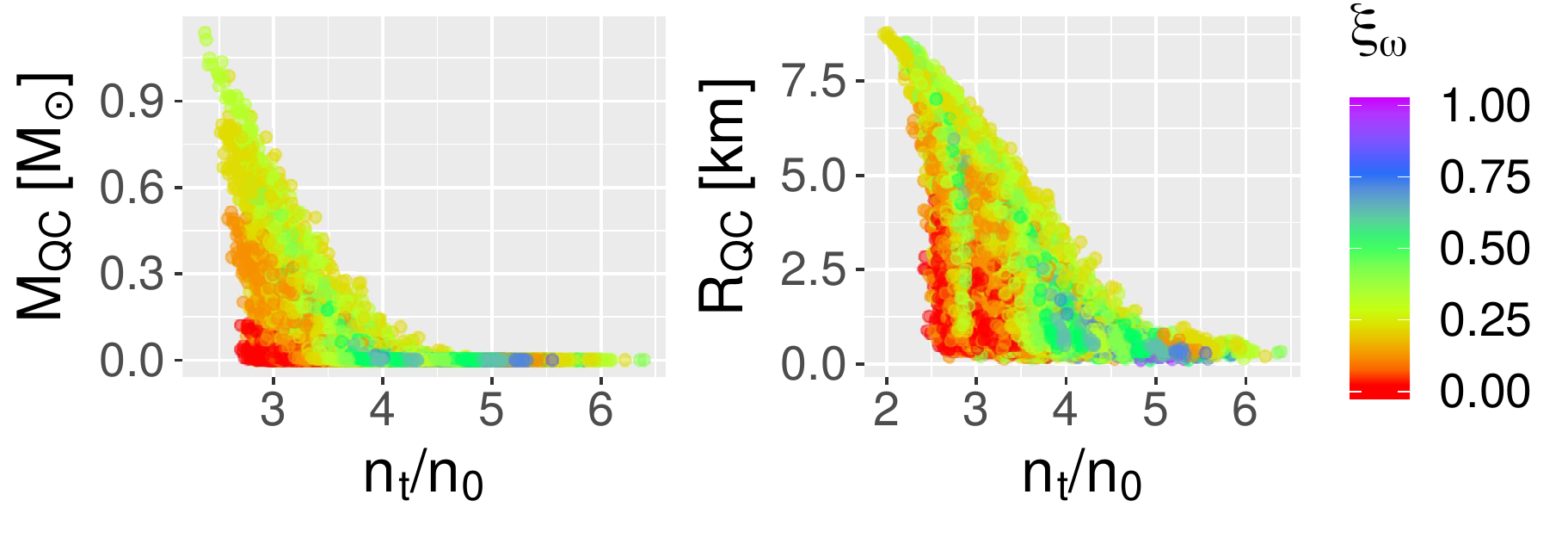}
	\caption{Size of the quark core as a function of the transition density $n_{t}/n_0$ 
		 for several values of $\xi_\omega$ (color scale).}
	\label{fig:result_3}
\end{figure}

Figure \ref{fig:result_3} shows  the core radius, $R_{QC}$, and core mass, $M_{QC}$, as a function of  the transition density $n_{t}/n_0$, 
where $n_{t}$ is the hadronic density at the phase transition, for several values of $\xi_\omega$.
As the value of $\xi_\omega=G_\omega / G_S$ increases, the quark matter becomes stiffer and the onset of quarks happens at larger
densities. On the other hand, quark cores become smaller as  $n_t$ increases.
Due the high dimensionality of the hadronic parametrization space and the computational cost of sampling a considerable number of hadronic models, we can only say that it is highly improbable of getting a valid hybrid EoS with $n_{t}<2.36n_0$. 
In the next section, we will focus on this point and, we will discuss the possibility of getting hybrid EoS with lower transition densities, and thus the existence of quark matter in light NS stars.  This strategy will reduce the computational cost and will 
 enable us to  explore a much larger hadronic EoS set.

\begin{figure}[!ht]
	\centering
	\includegraphics[draft=false,width=1.0\columnwidth]{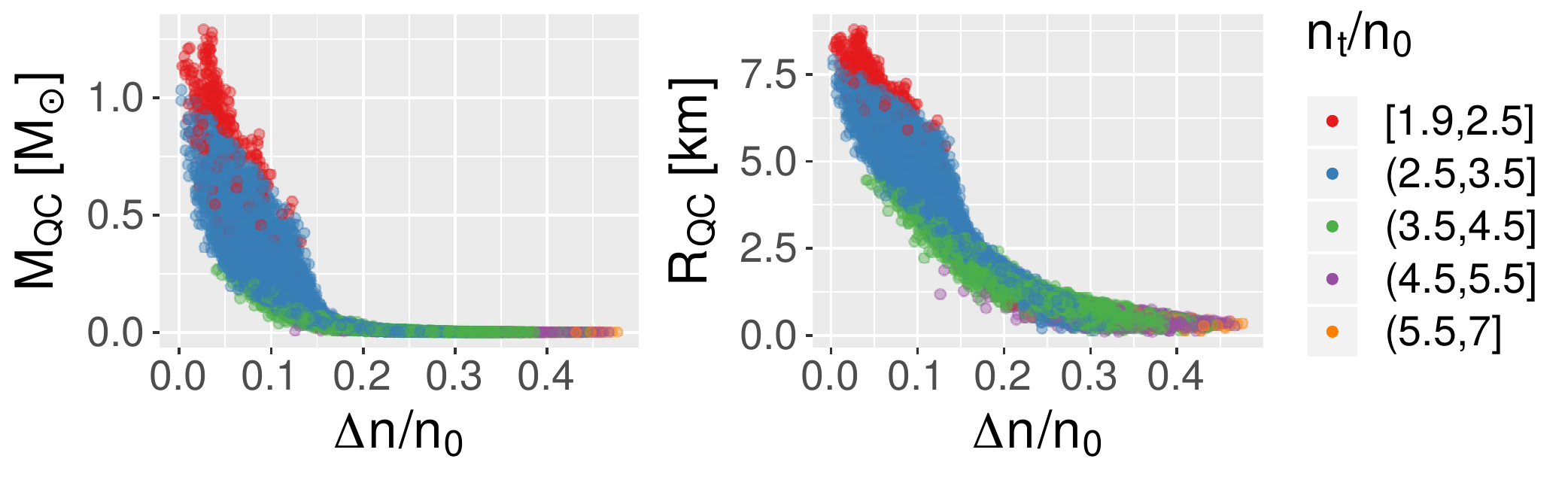}
	\caption{The quark core mass  $M_{QC}$ (left) and radius  $R_{QC}$ (right) as a function of transition density gap $\Delta n$ of each hybrid EoS. 
		On the color scale we show quark transition density $n_{t}$ value. }
	\label{fig:result_4}
\end{figure}

Next we analyze how the quark core depends on both the transition density $n_t$ and the density gap $\Delta n=n_q-n_{t}$ at the phase transition, where $n_{t}$ ($n_q$) is the baryon density in the hadronic phase (quark phase).
The results are in Fig. \ref{fig:result_4}. 
To sustain a considerable quark core size, the hybrid EoS must have both a small baryonic density gap $\Delta n$,
and it must occur for small $n_t$ values, $n_t<4n_0$. A large  $\Delta n$ and $n_t$ may still give a reasonably large  $R_{QC}$ corresponding to very stiff quark EoS, but the associated $M_{QC}$ are small. 
The required low $\Delta n$ and $n_t$ values to generate considerable quark cores, will be explored
when we study quark matter inside light NS.

\begin{figure}[!ht]
	\centering
	\includegraphics[draft=false,width=1.0\columnwidth]{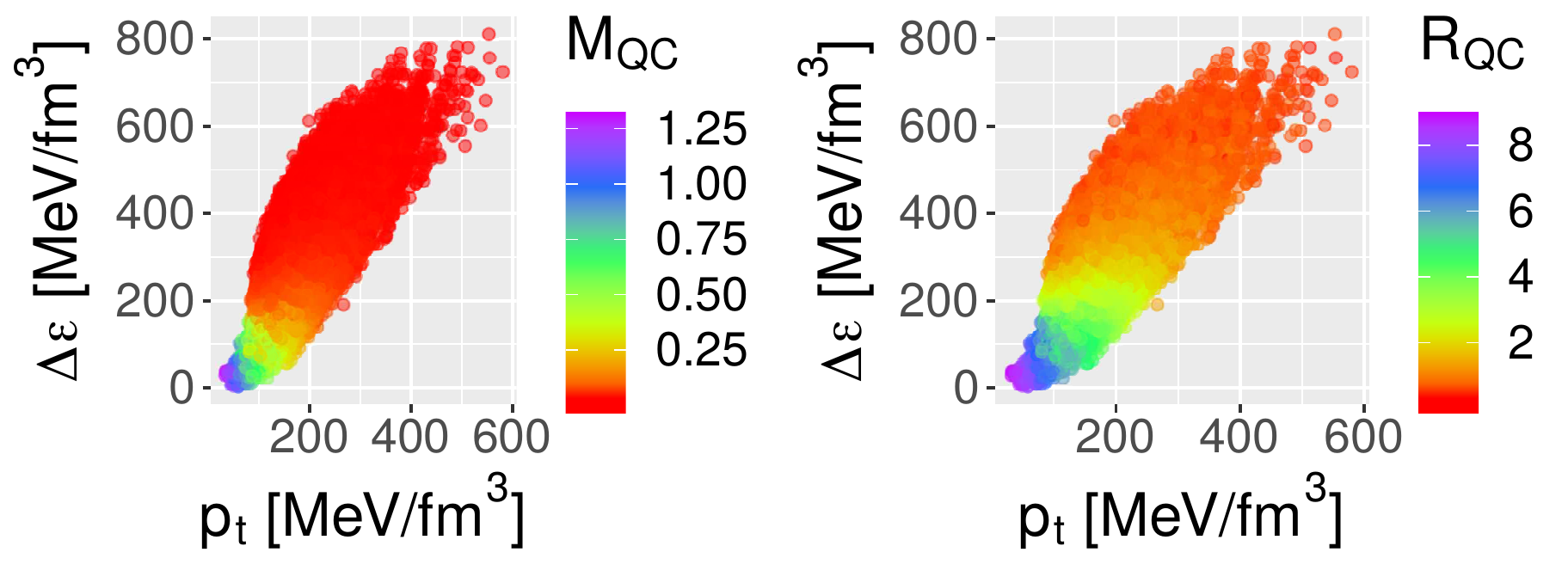}
	\caption{Dependence of the quark core size, $M_{QC}$ [$M_{\odot}$] (left) and $R_{QC}$ [km] (right), on both $\Delta$
	and $p_{t}$. }
	\label{fig:result_5}
\end{figure}

In Fig. \ref{fig:result_5}, we show how the quark core size depends on both the transition energy density gap, $\Delta \epsilon=\epsilon_q-\epsilon_{t}$, where $\epsilon_{t}$ ($\epsilon_q$) is the energy density in the hadronic (quark) phase at the transition, and the transition pressure $p_{t}$. 
As the phase transition sets in at low $p_{t}$ (low $n_{t}$), 
larger values of $M_{QC}$ and $R_{QC}$ can be sustained. 
To get a considerable quark core not only the energy gap must be small but also $p_{t}$, or, as we have already seen, $n_{t}$ must be small.
From Fig. \ref{fig:result_5}, we conclude that quite small quark cores, $M_{QC}\lesssim 0.25 M_{\odot}$, are obtained for $\Delta \epsilon>200$ MeV/fm$^3$ and $p_{t}>200$ MeV/fm$^3$.

The existence of a stable branch of stars with quark content depends on the
quark matter pressure being able to counterbalance the gravitational pull of the 
exterior hadronic mantle. An additional gravitational pull on the hadronic section of the star
raises from the phase transition energy density gap $\Delta \epsilon$, which is proportional to 
the $\Delta \epsilon$ value.
For high enough values of $\Delta \epsilon$, the quark core can no longer sustain 
the mantle gravitational pull and it becomes an unstable star right after the quark matter first appears.
Figure \ref{fig:result_5} shows precisely this: to generate a considerable quark core
both the amount of nuclear mantle must not be large, i.e, the phase transition must occur at low $p_{t}$,
and the $\Delta \epsilon$ cannot be large.
The precise values of both $\Delta \epsilon$ and $p_{t}$ depend on the properties of the quark EoS:
stiffer quark EoS are able to sustain higher values.

\begin{figure}[!ht]
	\centering
	\includegraphics[draft=false,width=1.0\columnwidth]{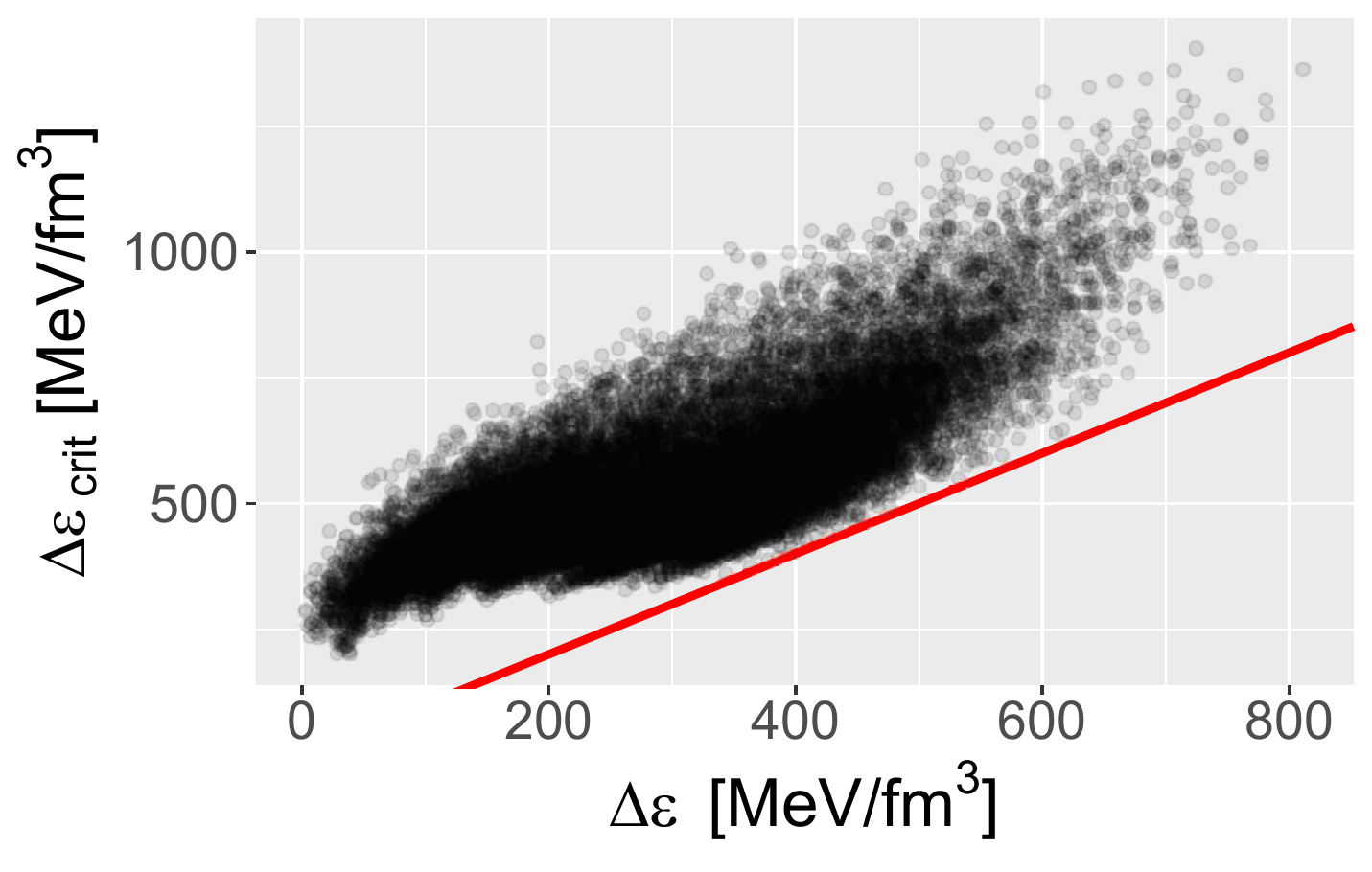}
	\caption{The $\Delta \epsilon$ versus $\Delta \epsilon_{crit}$ (see text) for each 
	hybrid star.}
	\label{fig:result_delta_crit}
\end{figure}

The existence of a quark core in a NS is determined by the hadron-quark phase transition properties \cite{seidov1971stability,schaeffer1983phase,Lindblom:1998dp}.
If the energy gap is beyond a critical value, $\Delta \epsilon > \Delta \epsilon_{crit}$, the presence of a quark core, regardless of its size, destabilizes the NS. 
The critical energy gap is given by $\Delta \epsilon_{crit}=e_{q}/2+3p_t/2$ \cite{Alford:2013aca}. 
In Fig. \ref{fig:result_delta_crit} we show $\Delta \epsilon$ vs. $\Delta \epsilon_{crit}$ for each hybrid EoS (the red line indicates $\Delta \epsilon=\Delta \epsilon_{crit}$).
Since our hybrid EoS are composed by quark branches that are connected to the hadronic branch,
 all our EoS satisfy $\Delta \epsilon<\Delta \epsilon_{crit}$ showing that they  indeed represent
stable hybrid stars solutions. 
A discussion of other types of solutions with $\Delta \epsilon<\Delta \epsilon_{crit}$, and
also $\Delta \epsilon>\Delta \epsilon_{crit}$, can be found in \cite{Alford:2013aca,Alford:2015gna}.

\begin{figure}[!ht]
	\centering
	\includegraphics[draft=false,width=0.8\columnwidth]{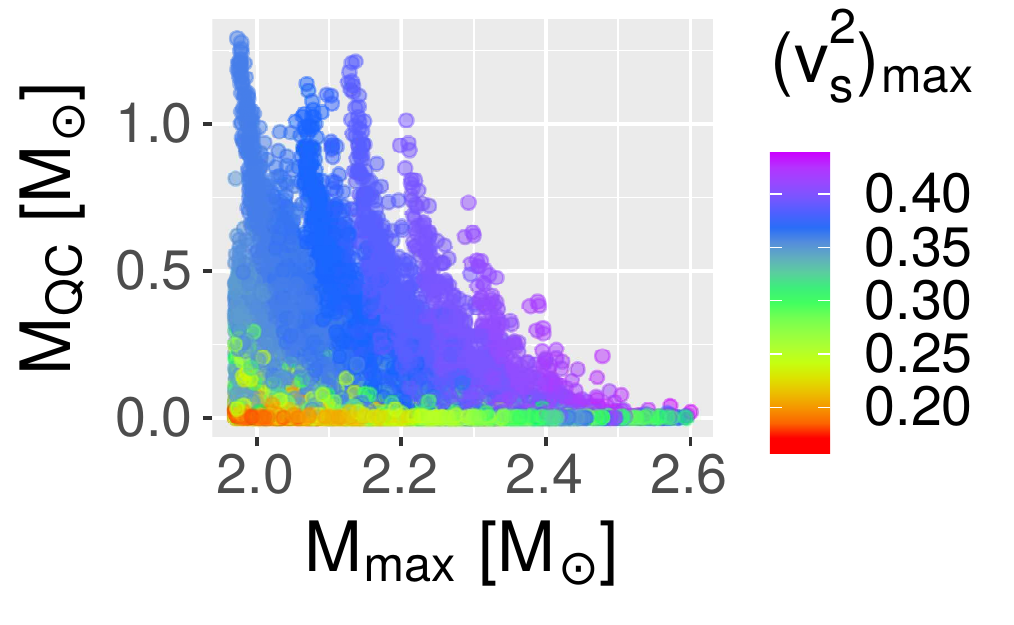}
	\caption{Core mass vs maximum mass as a function of the maximum values of $v_s^2$ (color scale) 
		reached in the quark phase for each hybrid star.}
	\label{fig:result_6}
\end{figure}

We have seen that the appearance of larger quark cores depends on the properties of the phase transition, namely
where the phase transition takes place, $p_{t}$ (or $n_{t}$), and on the size of its energy (or baryonic density) gap.
Larger cores are realized for low values of $p_t$ and $\Delta \epsilon$. However, the capacity for the quark core to sustain the hadronic mantle relies on the stiffness of the quark EoS. The parameter $\xi_\omega=G_\omega / G_S$ controls the stiffness of the quark EoS, and the higher its value the larger will be the quark core. Another way for analyzing the influence of the stiffness of the quark EoS is by looking at its sound speed, $v_s^2=dp/d\epsilon$.  In Fig. \ref{fig:result_6}, we show how $(M_{max},M_{QC})$ depends on the maximum value of $v_s^2$ reached in the quark phase for each hybrid EoS. 
Clearly, the quark EOSs that reach higher values for $v_s^2$ are able to sustain 
larger cores. The interpretation, as noted in \cite{Alford:2015gna}, is that a larger core can be sustained if its energy density rises slowly
enough compared with its pressure (making the sound speed higher), and a larger core with a higher central pressure may
be able to sustain the gravitational pull of the exterior hadronic mantle.
Therefore, if larger values of $v_s^2$  are reached by the quark matter then the energy
density of the core rises more slowly with increasing
pressure, which minimizes the tendency for a large core
to destabilize the star via its gravitational attraction (see \cite{Alford:2015gna}).

Within the NJL model, $v_s^2$ tends to $1/3$ for high enough values of $n$ as
it describes a phase of weakly interacting system of (almost) massless quarks.

\begin{figure}[!ht]
	\centering
	\includegraphics[draft=false,width=1.0\columnwidth]{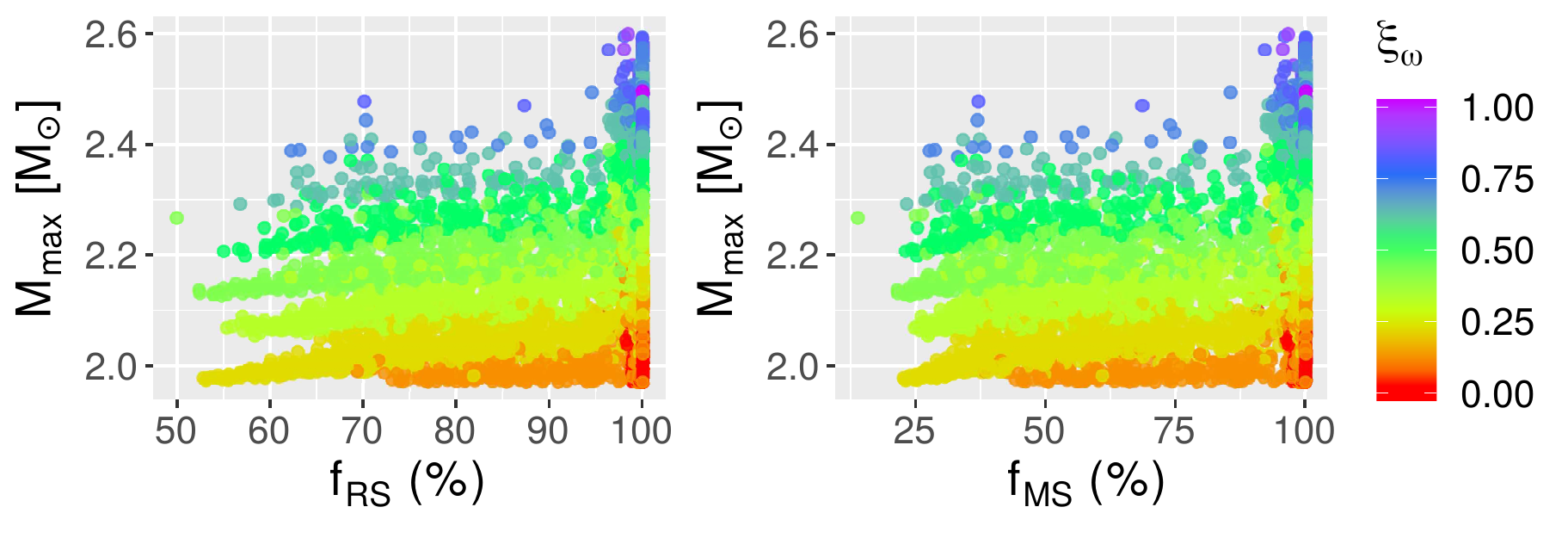}
	\caption{Maximum NS mass versus $f_{RS} = R_{SC}/R_{QC}$ for  the quark core radius (left) and  $f_{MS} = M_{SC}/M_{QC}$  for the quark core mass (right) (in \%) inside each hybrid star.
We show the value of $\xi_\omega$ in the color scale. }
	\label{fig:result_strangeness}
\end{figure}

We have discussed that the coupling $G_\rho$ has a direct influence on the onset of the strange quark. It is, therefore, of interest to analyze which is  the content of strangeness inside the quark cores of the hybrid stars under study.
We define the fraction of the quark core in which strange quarks are present as $f_{RS} = R_{SC}/R_{QC}$ and $f_{MS} = M_{SC}/M_{QC}$, where
$R_{SQ}$ ($M_{SQ}$) is the radius (mass) of the quark core with strange quarks. 
Figure \ref{fig:result_strangeness} shows both the strangeness fraction  $f_{RS}$ and  $f_{MS}$ on the radius and mass of the quark core.
Strange quarks are present in the whole core in the most massive NSs generated by each hybrid EoS. 
As $M_{max}$ increases, the quark core gets more strangeness content. This effect is easily  understood: in hybrid stars with a large maximum mass the quark phase must be very stiff and the phase transition from hadron to quark matter happens at densities where the strange quark already plays a role. As seen in \cite{Pereira:2016dfg}, increasing the vector interaction, for a fixed effective bag constant, increases the density at which the phase transition happens, pushing it to a part of the quark phase where the chemical potential is high enough to change down quarks into strange quarks.

\subsection{Hybrid EoS set with phase transitions between $1.3n_0$ and $2.5n_0$ \label{III3b}}

Now, we analyze the possibility of the presence of a quark core in light NSs by requiring that the onset of quarks, $n_t$, occur at  densities just above saturation density.
As we have shown, the NS mass in which quarks first appear, $M_t$, strongly depends on 
the phase transition properties. In order to generate a quark core for NS masses with $M\sim 1.4 M_\odot$
one needs smaller values of  density gap $\Delta n$ and energy density gap $\Delta \epsilon$, 
and the hadron-quark phase transition should occur at considerable low densities.
Thus keeping our lower bound of $0.2$ fm$^{-3}$, above which a phase transition is now allowed, 
we will also put an upper bound at $2.5n_0$:  the onset of a hadron-quark
phase transition is only  allowed in the density range $1.29<n_t/n_0<2.5$. 
Furthermore, the density gap should be such that $\Delta n/n_0<0.65$. 
These values can be justified by the analysis of Fig. \ref{fig:result_4}.
Considerable quark cores are obtained within the chosen range values for $\Delta n$ and $n_t$.
Defining a window where the phase transition could happen, instead of letting it occur at any density greater than $0.2$ fm$^{-3}$, 
as we did in the last section, allows us to sample a  larger number of hybrid EoS that predict light NSs with a quark core.

We have generated $5422$ hybrid EoS satisfying the above constraints. 
For this dataset, we consider a finer grid for the quark matter parametrization space:
 $\xi_\omega$ and $\xi_\rho \in [0,1]$ with $0.1$ intervals and $B\in [0,20]$ with $2$ MeV/fm$^{3}$ intervals.
The distribution of the transition densities is highly concentrated around the mean value $\bar{n_t}\approx  2.38 n_0$, 
being $n_t\approx1.84n_0$ the lowest transition density obtained. 
The hybrid EoS were constructed from a total of $595$ hadronic EoS,
whose statistics are in Table \ref{tab:hadron_2}.
The results in  Table \ref{tab:hadron_2} should be compared with 
Table \ref{tab:hadron}. The major differences are reflected in the iso-scalar properties of the EoS, in particular,  in the mean values 
of  the still unconstrained properties $Q_{sat}$ and $Z_{sat}$, and a slightly larger value of $K_{sat}$ but still
within the  established accepted range.  
The present set of hybrid EoSs is composed by hadronic matter with a specific range of 
empirical parameters that allow for a  first-order phase transition at low densities
to quark matter described by the NJL model. Note that the values in Table \ref{tab:hadron_2}
are still allowed due to our present uncertainty. 
All these hadronic EoS are quite stiff, sustaining heavy NS masses $M_{max}>2.48M_{\odot}$.
The phase transition must be weak, i.e.  with a low latent heat $\Delta \epsilon$,
enabling a not so stiff quark matter to sustain the nuclear mantle and generating stable sizable quark cores.

\begin{table}[ht]
	\centering
	\begin{tabular}{rrrrrrrrr}
		\hline
		 & $K_{sat}$ & $Q_{sat}$ & $Z_{sat}$ & $E_{sym}$ & $L_{sym}$ & $K_{sym}$ & $Q_{sym}$ & $Z_{sym}$ \\ 
		\hline
  mean & 244.55 & 434.37 & 557.83 & 33.38 & 49.90 & -18.07 & 195.39 & 400.46 \\ 
	std & 16.91 & 174.61 & 743.94 & 2.00 & 11.55 & 57.47 & 300.13 & 672.24 \\ 
		\hline
	\end{tabular}
\caption{Mean and standard deviation (Std) of the hadronic EoS. 
	All the quantities are in units of MeV. The total number of EoS is $595$.}
\label{tab:hadron_2}
\end{table}

\begin{figure}[!ht]
	\centering
	\includegraphics[draft=false,width=1.0\columnwidth]{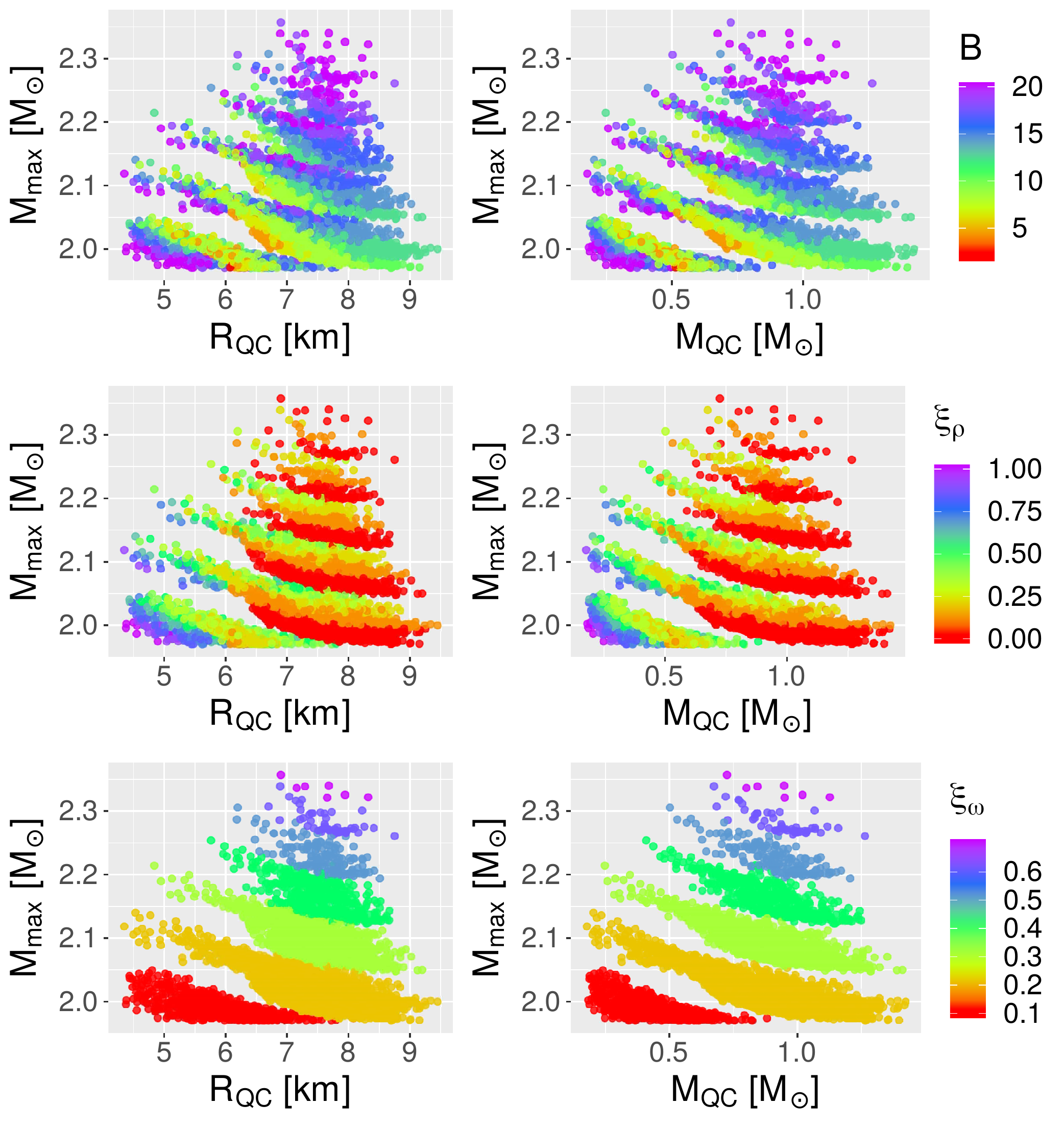}
	\caption{$M_{max}$ vs. $R_{QC}$ (left) and $M_{QC}$ (right) as a function of $B$ [MeV fm$^{-3}$] (top panels),
	$\xi_{\rho}$ (middle panels), and $\xi_{\omega}$ (bottom panels). }
	\label{fig:new1}
\end{figure}

Figure \ref{fig:new1} shows how the parameters $B$ (top panel), $\xi_{\rho}=G_\rho / G_S$ (middle panel), and $\xi_{\omega}=G_\omega / G_S$ (bottom panel) affect the radius and mass of the quark core inside the maximum mass configuration, represented in the
diagrams $\{M_{max}, R_{QC}\}$ (left) and $\{M_{max},M_{QC}\}$ (right). 
As noted before, the bands representing each value of $\xi_{\omega}$ (see in the bottom panels) show the strong impact it has on the relation 
between the quark core size and the maximum NS mass reached: the $M_{max}$ is proportional to the $\xi_{\omega}$ value. 
However, concerning the core size,  the biggest size is obtained with $\xi_{\omega}=0.2$, and for larger values the core size gets smaller.
Over each band with a specific $\xi_{\omega}$ value, there are multiple EoS with distinct $B$ and $\xi_{\rho}$ values.
Larger values of $\xi_{\rho}$ only describe hybrid EoS with lower $M_{max}$ values and small quark cores, 
i.e., to get considerable quark cores we must have  $\xi_{\rho}<0.3$, because a large $\xi_{\rho}$ softens a lot the quark EoS.

\begin{figure}[!ht]
	\centering
	\includegraphics[draft=false,width=1.0\columnwidth]{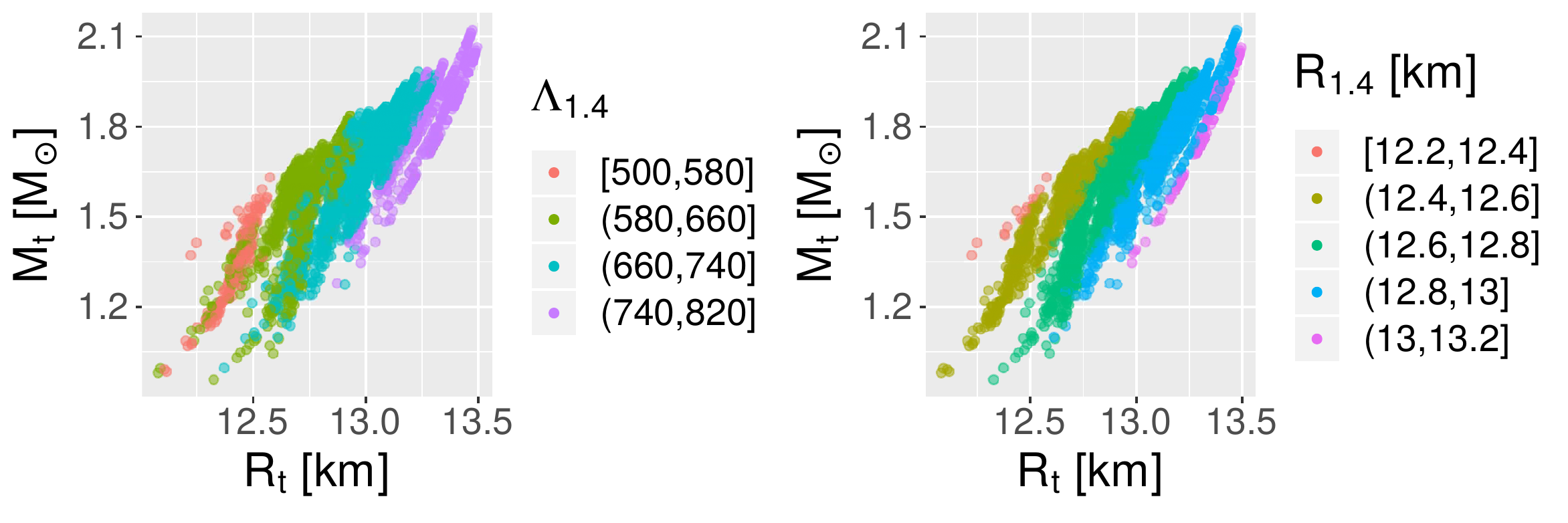}
	\caption{The NS mass and radius corresponding to the onset of quark matter, $M_t$ and $R_t$, for each hybrid EoS.
	The color scale shows the corresponding values of $\Lambda_{1.4M_{\odot}}$ (left) and $R_{1.4M_{\odot}}$ 
	(right) of each hybrid star. }
	\label{fig:new2}
\end{figure}

In Fig. \ref{fig:new2}, we show the mass $M_t$ and radius $R_t$ of the star where  quarks start to nucleate and the color scale indicates 
the corresponding values of $\Lambda_{1.4M_{\odot}}$ (left) and $R_{1.4M_{\odot}}$ for every hybrid EoS ($5422$ in total).
The first conclusion is that, in constraining the location where the phase transition must occur, 
we are able to have stars with quark content already at $0.96M_{\odot}$, not requiring the generation of a too large total number of EoS. 
We see that the  $\Lambda_{1.4M_{\odot}}$  is bound between $500$ and $820$. 
194 hybrid EoS of state fulfill the LIGO/Virgo constraint of  $\Lambda_{1.4M_{\odot}}<580$ \cite{Abbott18}, from which 
119 Hybrid EoS  predict a finite quark core for a $1.4M_{\odot}$ NS. 
Note, however, that this constraint on $\Lambda_{1.4M_{\odot}}$ obtained by the LIGO/Virgo collaboration does not require that the EoS used in the analysis support 1.97 $M_\odot$ stars.
The total of $5422$ hybrid EoS predict that $R_{1.4M_{\odot}}$ is in between $12.23$ km and $13.02$ km.
The subset of 194 hybrid EoS that satisfy $\Lambda_{1.4M_{\odot}}<580$ predicts $12.23<R_{1.4M_{\odot}}/\text{km}<12.57$ which are smaller than maximum radius predicted in \cite{Abbott18}, 12.8 km if the 1.97 $M_\odot$ constraint on the pulsar mass is not imposed, or 13.3 km otherwise.   These last constraints on the radius are satisfied by our complete set of  hybrid EoS. Our results show some friction between the current constraints on $\Lambda_{1.4M_{\odot}}$  and $R_{1.4M_{\odot}}$.
 The set of 119 Hybrid EoS that predict a quark core in a $1.4M_{\odot}$ NS and satisfy $\Lambda_{1.4M_{\odot}}<580$ is 
made of quark matter with 10 different hadronic parametrizations, with $B$ ranging from $10$ to $19$ MeV/fm$^3$, and
$\xi_{\omega}=0.1$, $0.2$ and $0.3$ and  $\xi_{\rho}=0$, $0.1$, $0.2$, and $0.3$.

\begin{figure}[!ht]
	\centering
	\includegraphics[draft=false,width=1.0\columnwidth]{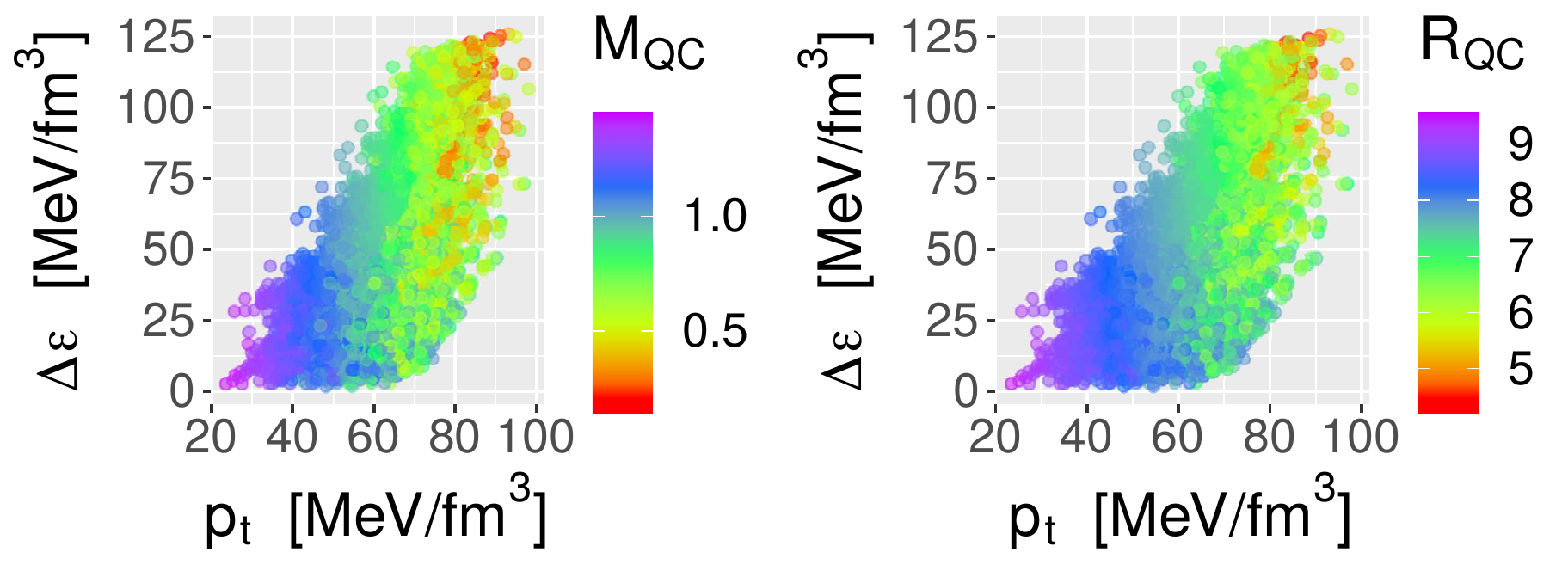}
	\caption{Diagrams $\Delta \epsilon$ vs. $p_t$ with the core size, $M_{QC}$ [$M_{\odot}$] (left) and $R_{QC}$ [km] (right), being displayed in the color scale.}
	\label{fig:result_9}
\end{figure}

Figure \ref{fig:result_9} shows the diagram  energy density gap $\Delta \epsilon$ versus $p_t$ and the color scale indicates the quark core size.
We see that the energy gap decreases as the transition pressure lowers. All these hybrid EoS correspond to weak first-order phase transitions from hadronic
to quark matter. The small latent heat of the phase transition, given by $\Delta \epsilon$, reflects this fact.
The values of $\Delta \epsilon$ range from $1.72$ MeV/fm$^{3}$ to $125.90$ MeV/fm$^{3}$, with a mean value 
of 49.80 MeV/fm$^{3}$. The $p_t$ has a mean value of $64.33$ MeV/fm$^{3}$, and minimum/maximum of $23.92/98.70$ MeV/fm$^{3}$.

\begin{figure}[H]
	\centering
	\includegraphics[draft=false,width=1.0\columnwidth]{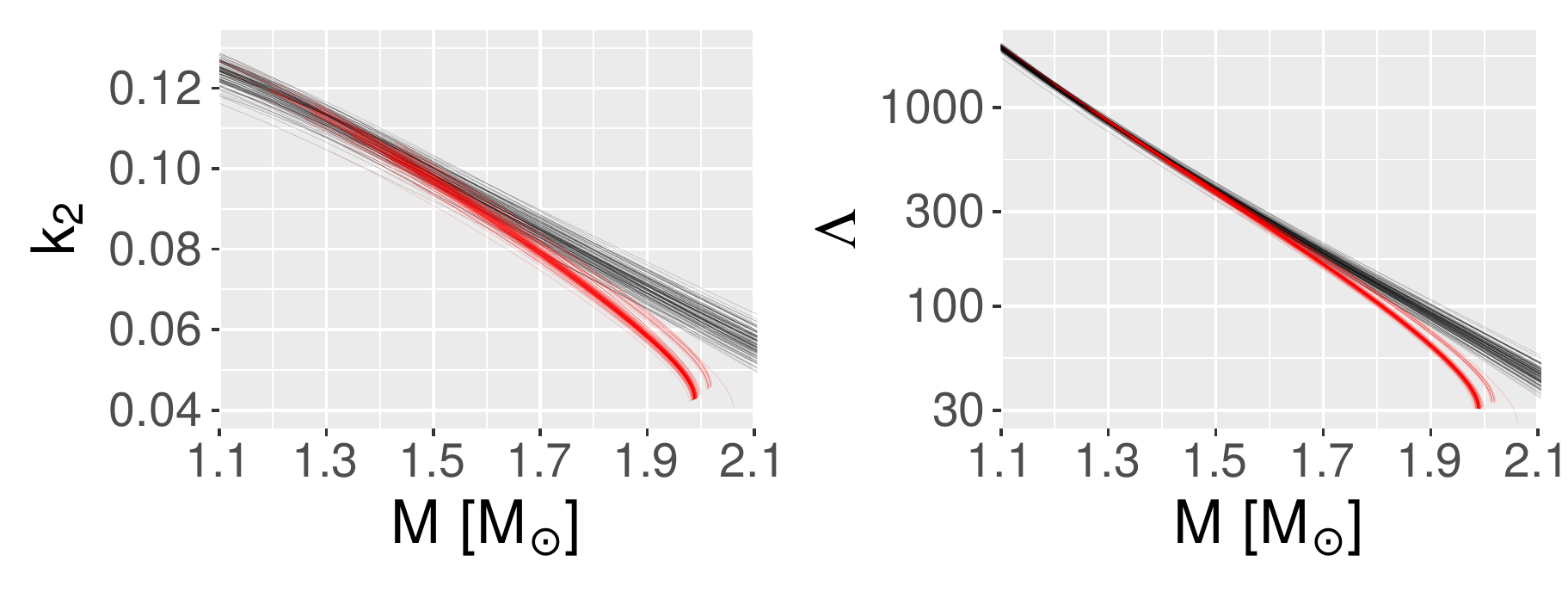}\\
	\includegraphics[draft=false,width=1.0\columnwidth]{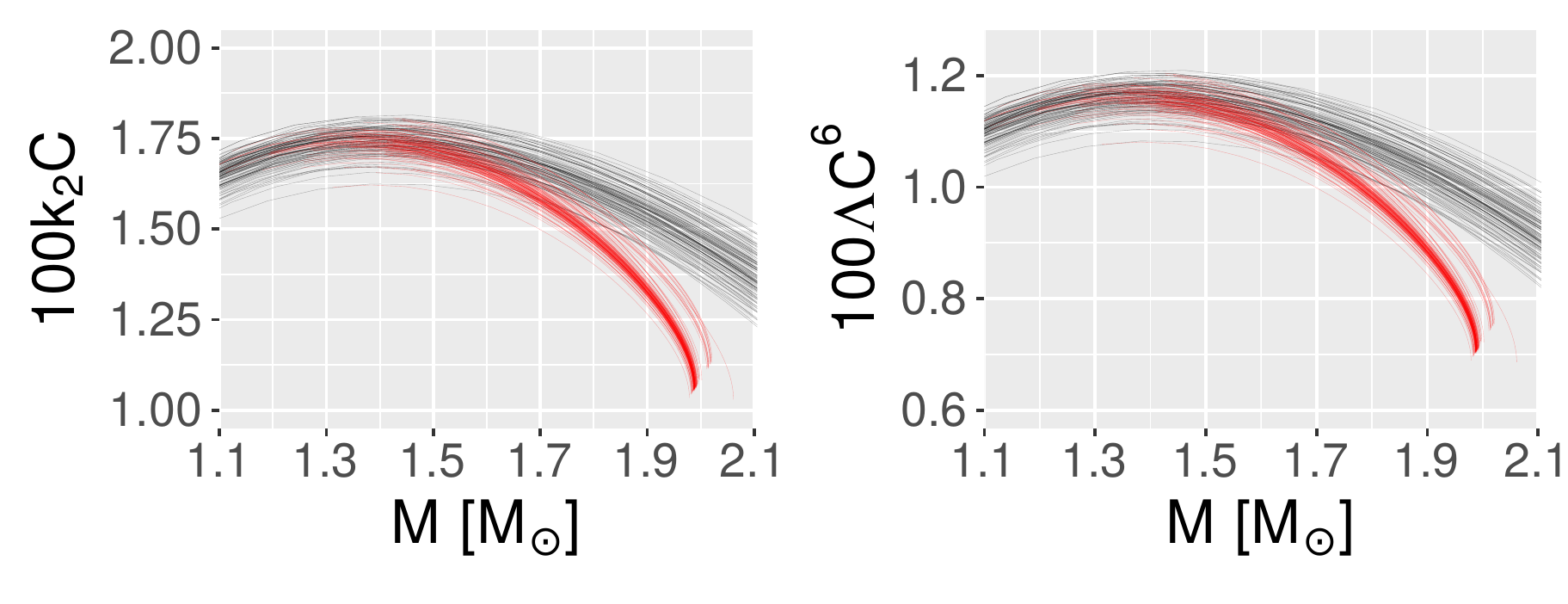}
	\caption{Love number $k_2$ and tidal deformability $\Lambda$ as a function of the NS mass for the 119 Hybrid EoS that predict
		 the existence of a quark core for $1.4M_{\odot}$ NS and $\Lambda_{1.4M_{\odot}}<580$.
	  In the bottom panel, we used the normalization factor of $100$.}
	\label{fig:result_10}
\end{figure}

Figure \ref{fig:result_10} shows $k_2(M)$ and $\Lambda(M)$ for the 119 Hybrid EoS that predict  the existence of a quark core 
for a $1.4M_{\odot}$ NS and that $\Lambda_{1.4M_{\odot}}<580$.
For comparison, we also show the result for purely hadronic matter. Despite the clear change in behavior 
when quark matter appears (red branches), it is a quite smooth change, reflecting the small $\Delta \epsilon$ of the phase transitions.
The relation between tidal deformability, $\Lambda$, and the tidal Love number, $k_2$, is given by 
$\Lambda = 2k_2C^{-5}/3$,
where $C=GM/(c^2R)$ is the star's compactness. 
In \cite{De:2018uhw}, the relation $\Lambda\sim C^{-6}$ was shown 
for hadronic stars with a mass in the range  $1.1\le M/M_\odot\le 1.6$.  The
extra $C^{-1}$ dependence comes from the tidal Love number $k_2$. 
In order to analyze the effect of a quark core in $\Lambda(C)$ and $k_2(C)$, we plot on the bottom panels $k_2C$  and $\Lambda C^6$ as a function of the NS mass. 
In the range $1.1\le M/M_\odot\le 1.6$, the variability of $\Lambda C^6$ is as large as $15\%$ and $18\%$ for the hadronic and hybrid EoS,
respectively. While for $1.1\le M/M_\odot\le 1.97$, $\Lambda C^6$ vary as large as $31\%$ for the hadronic and $63\%$ for he hybrid EoS.
The quark matter induces a different dependence on $k_2(C)$ and thus on $\Lambda$.
A clear deviation is seen on the tidal deformability for hybrid stars with $M>1.8\, M_\odot$.
From the figure it is also clear that because the quark matter sets in at low densities, and thus light NSs already have quark content,
these EoS are only able to sustain stars with maximum masses satisfying $1.98<M_{max}/M_{\odot}<2.06$.

\section{Conclusions}
\label{conclusions}

In this work we study the properties of hybrid neutron stars exploring a dataset of hybrid EoS built using a Taylor expansion for the hadronic phase and the NJL model for the quark phase. The hadron model parameters are related to empirical properties of nuclear matter at saturation and were sampled from an initial multi-dimensional Gaussian distribution. The hybrid stars were built considering a Maxwell phase transition from the hadronic to the quark phase, and requiring that each EoS from the set is monotonically increasing, causal, and support a maximum mass of at least $1.97M_{\odot}$. 
The NJL model used for the quark phase have three free parameters: the vector-isoscalar $\xi_\omega$ coupling, the vector-isovector $\xi_\rho$ coupling, and the effective bag constant $B$. 

The existence of a first-order phase transition influences the mass-radius curves: the sharp transition from hadron matter to quark matter leads to a connected branch of stable hybrid stars. The advantage of using a meta-model to build the hadron phase is that it gives flexibility at low densities to build hybrid EoS which fulfill the $1.97M_{\odot}$ maximum mass constraint while still providing empirical nuclear parameters that satisfy the present uncertainties on the EoS parameters, known from experiments and other theoretical studies. This flexibility in the hadron phase allows the existence  inside the NS of a core of quarks for a wide range of NJL parametrizations. 

Regarding the NJL parametrization, as already seen in previous studies, the effective bag constant plays an important role in building models which predict larger quark phases inside the star, since it allows to control the onset of the hadron-quark phase transition. Also, the presence of vector interactions is crucial to build hybrid stars with larger and more massive quark cores, specially the vector-isoscalar interaction ($\xi_\omega$ coupling) which makes the quark section of the hybrid EoS, stiffer. We have also observed that more massive stars have a very small quark core.

After the initial analysis restricting the onset of quark matter to happen at densities above 0.2fm$^{-3}$ we have realized that to sustain a considerable quark core size, the hybrid EoS must have both a small density gap $\Delta n$ at the transition, and it must occur at small $n_t$ values, $n_t<4n_0$. We have also verified that just a very small fraction of EoS predict quark matter for stars with $1.4M_{\odot}$. 

In order to study the possibility of the existence of a core with quark matter in light NS, we imposed restrictions in the constructions of the hybrid EoS, by requiring that the onset of quarks occurs at smaller densities. In this second data set, only a hadron-quark phase transitions in the density range $1.29<n_t/n_0<2.5$ were considered. Furthermore, the density gap was set to $\Delta n/n_0<0.65$. Both these requirements were motivated by the first study. Even though we were able to build such EoS which predict quark cores at low mass stars, they were not able to sustain very massive stars, and were limited to $1.98<M_{max}/M_{\odot}<2.06$.

The BNS mergers masses observed by the LIGO/Virgo collaborations are in the interval $1.0 \leq M / M_{\odot}\leq 1.8$. While the softening of the hybrid EoS, originated from the onset of quarks, might reduce the tidal deformability for NS masses around $1.4M_{\odot}$, in order to have quarks inside stars with such low masses, the hadron-quark phase transition must have a small energy gap. This means that, locally, there will not exist a big difference between the hadron phase and the quark phase, meaning it is very difficult to pinpoint a quark core observational signature that would imprint the low mass stars tidal deformability. 
There is a lot of tension between the observation of high massive pulsars, and the tidal deformability of the canonical star: the LIGO/Virgo data seems to indicate that the $1.4$M$_\odot$ has to be very compact, indicating that the EoS cannot be too hard, however the information coming from massive pulsars indicate that the EoS should be stiff in order to sustain at least $1.97$M$_\odot$. Building purely hadronic models that have such respect both these constraints is very challenging while such feature can be modeled by the hybrid EoS approach.

While it might be difficult to distinguish quark matter from hadronic matter through observational results on $(M,R)$, 
especially if the phase transition is weak,
other observable properties might indicate the presence of quark matter. 
The high neutrino emissivity of the quark core might lead to a fast cooling of the star or 
gravitational-wave emission of neutron star mergers may carry signatures of 
possible first-order hadron-quark phase transition at supra-nuclear densities \cite{Bauswein:2018bma}.

\section{Acknowledgments}

This work was partially supported by national funds from FCT (Funda\c c\~ao para a Ciência e a Tecnologia, I.P, Portugal) under the IDPASC Ph.D. program (International Doctorate Network in Particle Physics, Astrophysics and Cosmology), with the Grant No. PD/\-BD/128234/\-2016 (R.C.P.), and under the Projects No. UID/\-FIS/\-04564/\-2019, No. UID/\-04564/\-2020, and POCI-01-0145-FEDER-
029912 with financial support from POCI, in its FEDER
component, and by the FCT/MCTES budget through
national funds (OE).
The authors thank the Yukawa Institute for Theoretical Physics at Kyoto University, where this work was initiated during the workshop YITP-T-19-04 "Multi-Messenger Astrophysics in the Gravitational Wave Era".

\section{Thermodynamics of the NJL model}
\label{appendix}

In order to calculate the EoS of the NJL model, we consider the model in the so-called mean field approximation (MF). In this approximation, a quark bilinear operator, $\hat{ \mathcal{O} }$, is written as its mean field value plus a small perturbation, $\hat{ \mathcal{O} } = 
\expval*{ \mathcal{O} } + \delta\hat{ \mathcal{O} }$. Products of quark bilinears are then linearized by neglecting terms equal and superior to $(\delta\hat{ \mathcal{O} } )^2$. 

Using the Matsubara formalism \cite{Kapusta:2006pm}, the MF grand canonical potential for the SU(3) NJL model, at finite temperature and chemical potential, $\Omega$, can be written as: 
\begin{align*}
\Omega-\Omega_0 & = 
2 G_S  \qty(  \sigma_u^2 + \sigma_d^2 + \sigma_s^2 ) 
- 4 G_D\sigma_u\sigma_d \sigma_s 
\\
& -\frac{2}{3}G_\omega\left( \rho_u + \rho_d + \rho_s \right)^2  
\\
& - G_\rho\left( \rho_u - \rho_d  \right)^2 -
\frac{1}{3}G_\rho\left( \rho_u + \rho_d - 2\rho_s \right)^2
\\  
& - 2 N_c \sum_{i=u,d,s} 
\int \frac{ \dd[3]{p} }{(2\pi)^3}   
E_i 
\\  
& - 2 N_c T \sum_{i=u,d,s} 
\int \frac{ \dd[3]{p} }{(2\pi)^3}   
\ln \qty( 1 + \e^{-(E_i+\tilde{\mu}_i)/T} )
\\  
& - 2 N_c T \sum_{i=u,d,s} 
\int \frac{ \dd[3]{p} }{(2\pi)^3}   
\ln \qty( 1 + \e^{-(E_i-\tilde{\mu}_i)/T}  ) ,
\numberthis
\label{NJLpot}
\end{align*}
with $\Omega_0$ the value of the potential in the vacuum, $E_i=\sqrt{p^2+M_i^2}$, $\sigma_i$ and $\rho_i$ the $i$-quark flavor condensate and density, respectively. For $i\ne j\ne k\in \{u,d,s\}$, the effective mass is found to be given by,
\begin{align}
M_i = m_i 
- 4G_S \sigma_i
+ 2G_D \sigma_j \sigma_k ,
\end{align}
and the effective quark chemical potential by,
\begin{align}
\tilde{\mu}_i & = 
\mu_i 
- \frac{4}{3} 
[ 
\qty(G_\omega + 2G_\rho) \rho_i +
\qty(G_\omega -  G_\rho) \rho_j +
\qty(G_\omega -  G_\rho) \rho_k
] .
\end{align}
In the MF approximation the grand canonical potential must obey \cite{Buballa:2003qv},
\begin{align}
\pdv{\Omega}{M} = 
\pdv{\Omega}{\tilde{\mu}} = 0 .
\end{align}
Using these stationary conditions fixes the $i-$quark condensate and $i-$ quark density to be given by:
\begin{align*}
\sigma_i = 
\expval*{\bar{\psi}_i \psi_i} = &
- 2 N_c \int \frac{ \dd[3]{p} }{(2\pi)^3}  
\frac{M_i}{E_i} 
\\
&
+ 2 N_c \int \frac{ \dd[3]{p} }{(2\pi)^3}  
\frac{M_i}{E_i} 
\frac{1}{ \e^{(E_i-\tilde{\mu}_i)/T}+1 }  
\\
&
+ 2 N_c \int \frac{ \dd[3]{p} }{(2\pi)^3}  
\frac{M_i}{E_i} 
\frac{1}{ \e^{(E_i+\tilde{\mu}_i)/T}+1 } ,
\numberthis
\\
\rho_i =
2 N_c \int & \frac{ \dd[3]{p} }{(2\pi)^3} 
\qty[
\frac{1}{ \e^{(E_i-\tilde{\mu}_i)/T}+1 }  - 
\frac{1}{ \e^{(E_i+\tilde{\mu}_i)/T}+1 }
].
\numberthis
\end{align*}

Since we are interested in describing degenerate neutron star matter, we take the $T \to 0$ limit of the above equations.

The pressure and energy density of the system can be easily calculated from the grand canonical potential using:
\begin{align}
P & = -\Omega , 
\label{def.pressao}
\\
\epsilon & = -P +  \sum_i \mu_i \rho_i. 
\label{def.energia}
\end{align}

In the limit $T=0$, the quark condensate is given by:
\begin{align}
\sigma_i & = 
\expval*{\bar{\psi}_i \psi_i} = 
-\frac{N_c}{\pi^2} \int^{\Lambda}_{\lambda_{F_i}} \dd{p} p^2 \frac{M_i}{\sqrt{p^2+M_i^2}},
\end{align}
with the Fermi momentum, $\lambda_{F_i}=\sqrt{\tilde{\mu_i}^2 - M_i^2 }$ and quark density, $\rho_i = {N_c \lambda_{F_i}^3}/{3 \pi^2}$. The NJL pressure, in this limit is given by,
\begin{align*}
P  = & - \Omega_0 
- 2 G_S \qty( \sigma_u^2 + \sigma_d^2 + \sigma_s^2 ) 
+ 4 G_D \sigma_u \sigma_d \sigma_s  
\\
& + \frac{2}{3} G_\omega \left( \rho_u + \rho_d + \rho_s \right)^2 
\\
& + G_\rho \left( \rho_u - \rho_d \right)^2 
+ \frac{1}{3} G_\rho \left(  \rho_u + \rho_d - 2 \rho_s \right)^2
\\
&
+ \frac{N_c}{\pi^2} \sum_{i=u,d,s} \int^{\Lambda}_{\lambda_{F_i}} \dd{p}  p^2 E_i  + \frac{N_c}{\pi^2}\sum_{i=u,d,s} \tilde{\mu}_i \frac{\lambda_{F_i}^3}{3} ,
\numberthis
\end{align*}
while the energy density is,
\begin{align*}
\epsilon = & \Omega_0 
+ 2 G_S \qty( \sigma_u^2 + \sigma_d^2 + \sigma_s^2 ) 
- 4 G_D \sigma_u \sigma_d \sigma_s    
\\
&  - \frac{2}{3} G_\omega \left( \rho_u + \rho_d + \rho_s \right)^2 
\\
& - G_\rho \left( \rho_u - \rho_d \right)^2 
- \frac{1}{3} G_\rho \left(  \rho_u + \rho_d - 2 \rho_s \right)^2
\\
&
- \frac{N_c}{\pi^2} \sum_{i=u,d,s} \int^{\Lambda}_{\lambda_{F_i}} \dd{p}  p^2 E_i  + 
\frac{N_c}{\pi^2}\sum_{i=u,d,s} \left(\mu_i -\tilde{\mu}_i \right) \frac{\lambda_{F_i}^3}{3} .
\numberthis
\end{align*}

In order to study cold stellar matter, $\beta$-equilibrium and charge neutral matter must be imposed and a contribution coming from electrons, must be considered:
\begin{align*}
\Omega_e & = 
2T \int \frac{ \dd[3]{p} }{(2\pi)^3}  
\ln \qty( 1 + \e^{-(E_e+\mu_e)/T} )
\\
&
+
2T \int \frac{ \dd[3]{p} }{(2\pi)^3}   
\ln \qty( 1 + \e^{-(E_e-\mu_e)/T} ) ,
\numberthis
\label{potSU2_e}
\end{align*}
where $E_e=\sqrt{p^2+m_e^2}$ and $m_e=0.511$ MeV, is the electron mass.


%

\end{document}